\documentclass[table]{bmcart}

\usepackage{cite}
\usepackage{amsmath,amssymb,amsfonts}
\usepackage{graphicx,url}
\usepackage{textcomp}
\usepackage{float,booktabs}
\usepackage[ruled,linesnumbered]{algorithm2e}
\usepackage[group-separator={,}]{siunitx}
\usepackage{lipsum,multicol}
\usepackage[utf8]{inputenc}

\let\emptyset\varnothing
\usepackage[tight,footnotesize]{subfigure}
\usepackage{appendix}

\startlocaldefs
\endlocaldefs

\begin{document}

\begin{frontmatter}

\begin{fmbox}
\dochead{Research}


\title{Towards Business Partnership Recommendation Using User Opinion on Facebook}


\author[
   addressref={aff1},                   
   corref={aff1},                       
   email={diego.tsutsumi@alunos.utfpr.edu.br}   
]{\inits{DPT}\fnm{Diego P} \snm{Tsutsumi}}
\author[
   addressref={aff2},
   email={amanda.fenerich@pucpr.edu.br}
]{\inits{ATF}\fnm{Amanda T} \snm{Fenerich}}
\author[
   addressref={aff1},
   email={thiagoh@utfpr.edu.br}
]{\inits{THS}\fnm{Thiago H} \snm{Silva}}


\address[id=aff1]{
  \orgname{Department of Informatics, Federal University of Technology - Parana}, 
  \street{Sete de Setembro Avenue, 3165},                     %
  \city{Curitiba},                              
  \cny{Brazil}                                    
}
\address[id=aff2]{
  \orgname{Graduate Program in Industrial Engineering and Systems, PUC - Parana}, 
  \street{Imaculada Conceição St., 1155},                     %
  \city{Curitiba},                              
  \cny{Brazil}                                    
}


\begin{artnotes}
\end{artnotes}

\end{fmbox}


\begin{abstractbox}

\begin{abstract} 
The identification of strategic business partnerships can potentially provide
competitive advantages for businesses; however, \textcolor{black}{ due to the dynamics and uncertainty present in business environments }, this task could be challenging. To help \textcolor{black}{ businesses in this task }, this study presents a similarity model between businesses that consider the opinions of users on content shared by businesses on social media. Thus, this model captures significant virtual relationships among businesses that are generated by users in the virtual world. Besides, we propose an algorithm for detecting business communities in the considered model. We also propose an algorithm to identify possible business outliers in the detected communities, which could represent an automatic way to identify non-obvious relations that might deserve particular attention of business owners. By exploring approximately 280 million user reactions on Facebook, we show that our results could favor the development of, for example, a new strategic business partnership recommendation service.
\end{abstract}




\end{abstractbox}
%

\end{frontmatter}




\section{Introduction}

Strategic business partnerships are essential for various reasons. For instance, it could favor a competitive advantage for the business. A partnership with a true win-win intention could provide the edge a business needs to surpass its competitors. However, a poorly thought out partnership can hinder instead of help, making this procedure challenging \cite{bergquist1995building,elmuti2001overview}.

Recently, businesses from different segments have been exploring social media for various purposes, including for marketing. That involves producing and sharing content on social media platforms to promote a service or product, envisioning to achieve branding goals \cite{hoffman2010can,tuten2017social}. Thus, social media platforms become also crucial for customer relationship management, among other things \cite{ferrari2016content,chaffey2016global}.

Those interactions in social media generate a considerable amount of customer-business relationship data. Exploring these data could be an interesting alternative to complement traditional business analysis, such as market analysis and market segmentation, which, typically, do not scale easily.  Since the collection of customer-business relationship data on social media can be cheaper and faster, their proper analysis enables a market study in possibly shorter time and fewer effort \cite{culotta2016mining}.

We know that many factors, such as tastes and opinions, could affect the customers' preferences for business \cite{trainor2014social,agnihotri2016social,hudson2013impact}. We believe that users' preferences for specific business could be implicitly manifested in the actions performed in the content shared by those companies in social media. In some social media platforms, such as Facebook, users can react to the companies content. Thus, these reactions could be a proxy to capture these implicit preferences. Preferences implicitly manifested by users in actions in social media was also assumed to exist in previous works \cite{cranshaw2012livehoods,silva2014you,Mueller2017,brito18}.

To contribute to the \textcolor{black}{ task } of identifying strategic business partnership, this study aims to find significant virtual relationships among businesses that are generated spontaneously by users in social media. The first step towards that is by proposing a similarity model between businesses that consider the opinions, i.e., reactions, of users on content shared by businesses on social media. For that, it is used public data on social media. Particularly, we explored more than 280 million public user reactions collected from Facebook's users about businesses in Curitiba, PR, Brazil. Facebook, among many other social media platforms, has been chosen because it is a popular social media among customers and companies \cite{ferrari2016content}. 

Besides, we also provide an iterative algorithm for detecting businesses communities in the presented model. \textcolor{black}{ A community can be viewed as a cohesive and densely connected subgraph in a larger graph structure; furthermore, algorithms for community detection in graph structures are well known in the literature \cite{fortunato2010community,rossetti2018community}. In this study, a specific domain algorithm for business community detection is proposed taking into account all the background in the literature. }


We found that it enables the identification of business communities that have a surprising similarity regarding categories of businesses inside each community.  Even not considering any information of the businesses themselves, all communities have strong semantic similarities in business that compose them, indicating that our approach has the potential to extract cohesive communities of business. We also proposed an automatic approach to extract possible relations outliers, business, on those communities. We observe that non-obvious relationships between businesses could be extracted by using that approach, which might deserve particular attention of business owners.

Our results favor the development of new services and applications. For example, a new strategic business partnership recommendation service that can be useful for entrepreneurs and business owners. This service can contribute to sales improvement, as well as to keep companies more sustainable in the market.

We organize the rest of the study as follows. Section \ref{sec:relatedWork} presents some of the main related work to this study. Section \ref{sec:datacollection} describes the particularities of the data collected, which are used in the model proposed in Section \ref{sec:modelanalysis}. Section \ref{sec:results} presents the results obtained by this study. Finally, Section \ref{sec:conclusion} concludes the paper and points out future work.

\section{Related Work}
\label{sec:relatedWork}


According to Mukhopadhyay \cite{mukhopadhyay2018opinion}, with the advent of online technologies and social media, individuals are increasingly sharing their views and opinions through the internet, which are influencing and affecting the business sociopolitical and personal contexts. A considerable amount of effort has been made by the academic community with the objective of extracting relevant information from social media, which is evidenced by the constant growth of literature. For instance, the average annual growth rate of the number of publications over the last five years is around $0.15$ at Scopus and $0.37$ at Web of Science Core Collection\footnote{These values represent average growth using the term ``Social Media'', and ``Social Media'' and ``Business'' in conjunction.}, and a similar trend is observed in other scientific databases. 

The information from social media can be used for different purposes \cite{SilvaCSUR2019}, ranging from mobility understanding \cite{hudson2013impact,anaBeyondSights} and well-being improvement \cite{schwartz2013characterizing,DeChoudhury2016} to city semantics understanding \cite{aiello2016chatty,santosWI2018} and gender behavior study \cite{Mueller2017,Magno2014}. Some of the studies in this direction have implications for existing and new business. 

For instance, Cheng et al. \cite{cheng2011exploring} presented spatiotemporal analyses of users' displacement, exploring for that a dataset from a Foursquare-like system, i.e., a social media for location sharing. Their results can provide support in decisions about where and when to invest resources in a new business. By observing what people eat and drink in location-based social networks, Silva et al. \cite{silva2014you} developed an approach to identify boundaries between different cultures, which could be useful, for example, to businesses from a particular country that desire to verify the compatibility of cultural preferences across different markets.

Cranshaw et al. \cite{cranshaw2012livehoods} introduced the concept of Livehoods, which are regions of a city partitioned and grouped by similarity of users' behaviors. This behavioral targeting study may also be necessary for strategic decisions in companies. The geographic interaction characteristics of online public opinion propagation were also studied by Ai et al. \cite{ai2017national}. Barbier et al. \cite{barbier2011understanding} proposed a method to understand the behavior and dynamics of online groups, showing that their results could have practical business implications, such as a better understanding of customers and influence propagation. 

Yang et al. \cite{yang2017structural} presented the concept of core-periphery structures, that is a two-class partition of nodes (one is the core, and the other is the periphery), and provide empirical evidence that social communities always have this type of structure. This is an important study in the classic task of community detection, in social network analysis, which has implications on the different influences among users within the network. Thus, the core-periphery structures enable identification of such an influential actor, a business in our case, in a community. \textcolor{black}{ Regarding community detection, Fortunato \cite{fortunato2010community} formally defines communities and presents a comparison of some algorithms for detecting communities. Rossetti and Gazabet \cite{rossetti2018community} also give a background on community discovery; however, in contrast to \cite{fortunato2010community}, they state community structures in dynamic networks. }

Wu et al. \cite{wu2016simple} were also concerned with studying a community detection task, presenting a new framework for detecting parallel communities, called SIMPLifying and Ensembling (SIMPLE). Similarly, Liu et al. \cite{liu2017markov} used the Markov-network for discovering latent links, i.e., links which are not directly observable but rather inferred, among people in social networks. Data analysis of social media based on community detection was also used by Alamsyah et al. \cite{alamsyah2017social}, and some challenges of community detection in social media were presented by Tang and Liu \cite{tang2010understanding}. Social community detection studies are important because they can be applied to networks of different types. In fact, this study presents a business network demonstrating the utility of social community detection in the business context.

Grizane and Jurgelane \cite{grizane2017social} reinforce the importance of social media as a marketing tool for business, and present a model assessing the benefits of investing financial resources in social media.  Mahony et al. \cite{mahony2018if} investigate the adoption of social media by small business, showing that the use of social media can bring benefits not only to large companies but also to small and medium businesses. Kafeza et al. \cite{kafeza2017exploiting} also focused on assessing business process performance using social media analysis and community detection methods, but with the differential of examining how communities change in time. 

An interactive community mapping and detection scheme to reveal the dynamics of communities' evolution around an event is proposed by Giatsoglou et al. \cite{giatsoglou2015user}. This approach to identifying static and evolutionary communities over time is of particular interest in investigating business dynamism. Dynamic evolution over time was also studied by Pepin et al. \cite{pepin2017visual}, who presented an approach based on a graph model easily adaptable to an interactive visualization. Visual analysis can be useful, for example, in the presentation of business communities and how they change over the years.

Considering that the contents of social media are dynamic, Palsetia et al. \cite{palsetia2014excavating} presented an approach for detecting social communities based on posts, comments, and tweets of users. Their approach is, in some aspects, similar to Algorithm \ref{alg:businessCommunity} presented in this study, they iteratively remove communities from the main graph making the detection of the current community not influenced by previously detected communities.

Closer to the proposal of this study, two studies were developed to help entrepreneurs and decision makers to find the best place in a city to open a new business, and the core decision process is based on social media data \cite{lin2016goldmine,karamshuk2013geo}. Regarding the study of Lin et al. \cite{lin2016goldmine}, business information, such as business type, location, and check-ins, is collected from public pages of Facebook in order to recommend the best places in the city of Singapore to open a new business. Similarly, Karamshuk et al. \cite{karamshuk2013geo} collected information from Foursquare also aiming to recommend better locations to business, but in contrast to the static data explored in the study of Lin, Karamshuk also considered users' mobility.

The authors of this present study have previously performed a study in this direction \cite{diegoCourb18}. To the best of our knowledge, \cite{diegoCourb18} differs from all other studies available in the literature, since it aims to identify virtual relationships among businesses that are generated spontaneously by users in social media. To achieve this goal, that study proposes a new way of modeling these relationships, as well as a strategy to extract relevant connections among businesses. Also, it presents a strategy for detecting businesses communities in the proposed model. The present study significantly builds upon our previous work \cite{diegoCourb18}. First, it is proposed in this study a new approach to extract relations outliers on the communities detected. \textcolor{black}{ From the outliers }, we observe that non-obvious relationships between businesses could be extracted by using that approach, improving the discussion and analysis of communities detected. Also, it is presented important properties of our dataset, and it is discussed more details about the proposed model, for instance, key statistics of the model.

It is important to note that our study could be used in conjunction with some of the previous efforts. For instance, the model proposed by Grizane and Jurgelane \cite{grizane2017social} could be used in conjunction with the model proposed in this research (explained in Section \ref{sec:modelanalysis}),  in order to enrich the information provided to entrepreneurs about the impacts of the use of social media in business.

\section{Data Collection and Processing}
\label{sec:datacollection}
\subsection{Data Choice}
\label{sec:DataChoice}

The decision of what data to collect is important to support further analysis, as well as to meet possible limitations imposed in the data collection process. For this study, data were collected from Facebook, \textcolor{black}{ because it is the most used social media platform in Brazil \cite{ferrari2016content}. According to Ferrari et al. \cite{ferrari2016content}, in Brazil alone the number of Facebook users reached 74.8 million. Facebook is also highly relevant for businesses to create new and maintain ongoing relationships with their customers; therefore, data are widely available for analysis \cite{ferrari2016content}. }

As the purpose of the model is to identify virtual relationships among businesses exploring user reaction data on Facebook, the data were chosen considering our objectives and what is publicly available on Facebook. Table \ref{tab:datastructure} displays the structure of the considered data. We could choose similar information from other social media platforms; however, this assessment is outside the scope of this present study.

\begin{table}[ht]
    \centering
    \caption{Structure of the considered data.}
    \label{tab:datastructure}
    \begin{tabular}{|ll|p{2.5cm}l|}
    \toprule
\textbf{Business Data} & Example & \textbf{User Reaction Data} & Example\\ \midrule
Business ID & 166765230043005 & User ID & 154573625550 \\
Business Name & Rubiane Frutos do Mar & User Name & Fulano de Tal\\
Location & -25.516122, 49.231571 & Business ID that User Reacted to & 166765230043005\\
Category & Seafood Restaurant & Reaction Type & \textit{Like} \\
Number of Check-ins & 38627 & & \\
Number of Fans & 15532 & & \\
Average Evaluation & 4.6 & & \\ \bottomrule
\end{tabular}
\end{table}

The first column of Table \ref{tab:datastructure}, called \textit{Business Data}, represents data referring to the businesses themselves, such as their geographical location, their category (i.e., the market sector in which the business operates) and so forth. Therefore, each business in the dataset has all the information described in the first column. The second column contains \textit{User Reaction Data} for businesses in our dataset. Each reaction expressed on Facebook comes from a user and refers to a particular business. 

User reaction data makes it possible to create similarity connections among businesses, primarily by using the common reactions expressed by users regarding two or more businesses, as explained in Section \ref{sec:modelanalysis}. The business data is used mainly to extract names and categories of the businesses, which assists the reaction collection and evaluation of the results presented in Section \ref{sec:results}.

All information collected is open and publicly available on the Facebook platform. More details can be found on the Facebook Graph API\footnote{https://developers.facebook.com.}.

\subsection{Data Collection}

\begin{figure}[ht]
    \centering
    \includegraphics[width=0.95\textwidth]{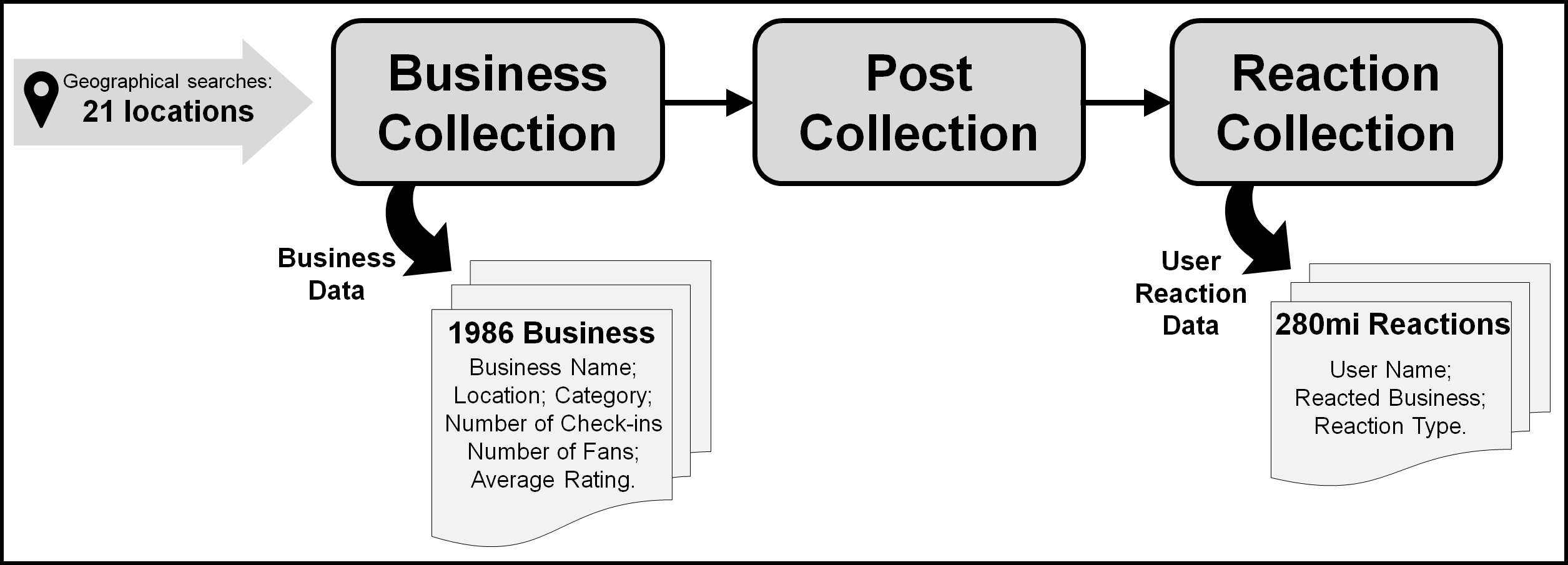}
    \caption{Illustration of the data collection process.}
    \label{fig:DataCollMethod}
\end{figure}

\begin{figure}[ht]
    \centering
    \begin{minipage}{0.47\linewidth}
        \includegraphics[width=\textwidth]{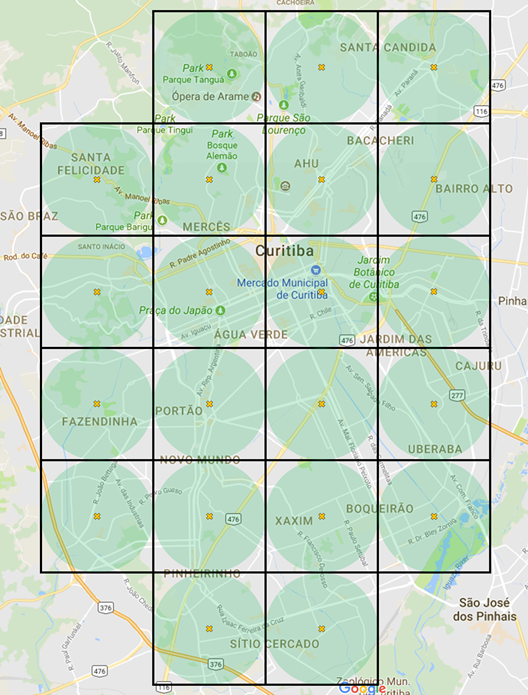}
        \caption{Data collection points considered for Curitiba, PR, Brazil.}
        \label{fig:DataCollGrid}
    \end{minipage}%
    \hfill
    \begin{minipage}{0.47\linewidth}
        \includegraphics[width=\textwidth]{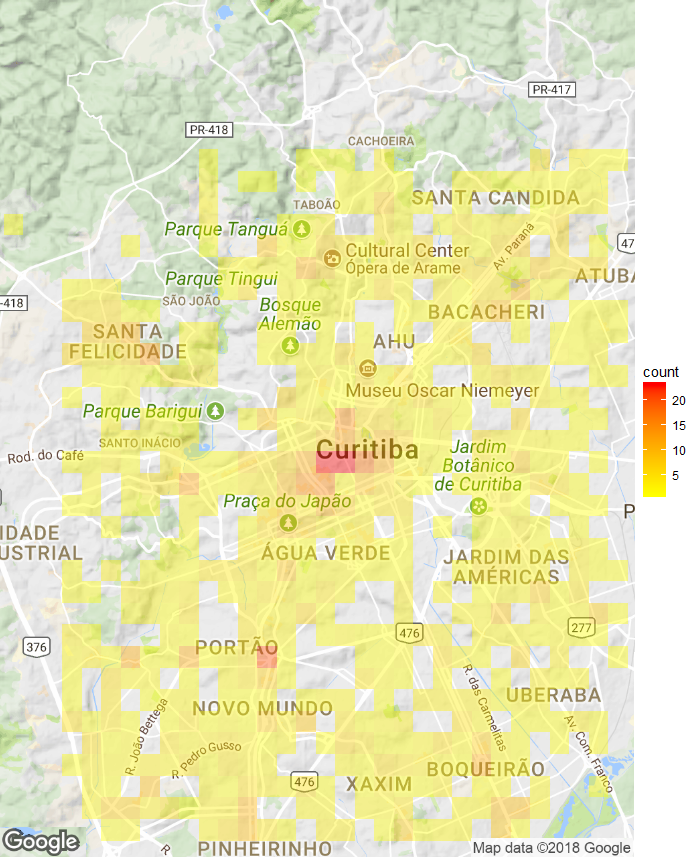}
        \caption{Heatmap representing the number of businesses found in different regions by the business search process.}
        \label{fig:HeatMap}
    \end{minipage}%
\end{figure}

Figure \ref{fig:DataCollMethod} illustrates the main steps performed in the collection process. The data were collected using the Facebook Graph API\footnote{https://developers.facebook.com/docs/graph-api/overview.}, all business data collected are located in the city of Curitiba, Brazil, and all user reaction data are from November to December of 2017. First, we collected data referring to the first column of Table \ref{tab:datastructure}, i.e., \textit{Business Data}. Facebook Graph API requires a geographical coordinate and a radius in meters, then returns results considering the coordinate entered as the center of a circle with the radius informed, returning up to $800$ results per search, that is, up to $ 800 $ businesses in this case. It is known that several regions of the studied city may have this number of businesses, then to increase the chance of getting most businesses of all regions of Curitiba, we considered twenty-one different geographic searches throughout the city. Each geographic search has a radius of $ 2000 $ meters and is centered in different regions of Curitiba, as shown in Figure \ref{fig:DataCollGrid}.

Figure \ref{fig:HeatMap} shows a heatmap representing the number of business found in different regions by the collection process just described. The redder the color, the more business were found in that particular location. As we can see, the central region of the city has a reddish coloration, indicating that more businesses were collected in that region, as expected. Despite that, it is also possible to notice that the resulting dataset includes businesses spread all around the city of Curitiba. Figure \ref{fig:mapLike} in Appendix \ref{app:reacDatabase} shows how user reactions are distributed across regions of the city, as expected, it is easy to notice that the amount of reactions is more significant in the city center.

After obtaining the results of the geographical searches (\textit{Business Data} in Table \ref{tab:datastructure}), containing basic data of the businesses in Curitiba, the reactions of the users (\textit{User Reaction Data} in the Table \ref{tab:datastructure}) were collected from the business pages previously collected. For that, we obtained the reactions of the posts on the business pages. Because some businesses' pages have hundreds of posts and others have millions, we only collected reactions of the first one hundred posts. There are five types of reactions available in Facebook, namely \textit{Like, Angry, Wow, Sad and Thankful}; we included all types in the database.

We collected a total of $1,986$ georeferenced pages and approximately $280$ million user reactions related to those pages. In Appendix \ref{app:reacDatabase} it is presented supplementary information about this dataset. Figure \ref{fig:compLike}, shows the top twenty businesses regarding user reaction number, in addition, Figure \ref{fig:catLike} shows the top twenty businesses categories concerning user reaction number.

\subsection{Data cleaning}

After obtaining all the data, an automatic \textcolor{black}{ and manual } cleaning procedure was performed to increase the consistency of the dataset, consequently increasing the consistency of the final results. We performed three main steps:

\begin{itemize}
    \item Duplicate records removal \textcolor{black}{ (automatic procedure) };
    \item Inconsistent records (e.g., unnamed pages, without location, and) removal \textcolor{black}{ (automatic procedure) };
    \item Removal of pages that do not represent businesses (for example, a public square) and their reactions \textcolor{black}{ (automatic and manual procedure) }.
\end{itemize}

After the cleaning process, the dataset is left with $1,926$ pages, all representing businesses, and approximately $260$ million reactions.

\section{Modeling and Strategies for Data Analysis}
\label{sec:modelanalysis}

\subsection{Overview}

\begin{figure}[ht]
    \centering
    \includegraphics[width=0.95\textwidth]{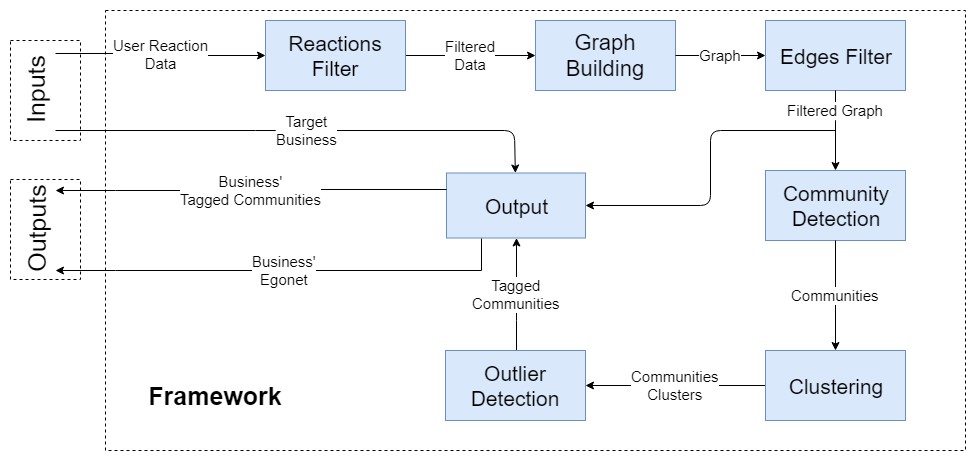}
    \caption{General view of the proposed analysis framework for identifying virtual relations among businesses with social media data.}
    \label{fig:Fluxograma}
\end{figure}

The main steps employed in this study to achieve the proposed objectives can be described in a framework\textcolor{black}{ , proposed by the authors of this study, } illustrated in Figure \ref{fig:Fluxograma}. The framework entries are the \textit{Clean Data}, described in Section \ref{sec:datacollection}, and a \textit{Target Business}, the business chosen to be analyzed. As outputs, we obtain the \textit{Egonet of the Target Business}, a network of the most relevant direct connections of the target business, and the \textit{Business Tagged Communities}, tagged communities of businesses in which the target business is part of. All non-standard businesses inside communities, considered outliers, are tagged. We provide more details about those outputs next. \textcolor{black}{ A relevant feature of this framework is that it is designed to handle data from any data source. In this paper it is used Facebook data; however, this is not a restriction of the framework. }


\subsection{Business Relationship Graph} \label{sec:graph}

With the obtained data, we can then create a model to represent the virtual relations among businesses. The model is a non-directed graph in which vertices represent businesses, and weighted edges represent relations between two businesses. This relationship is built looking at the reactions of users in common between any two businesses. The more common reactions two businesses have in proportion to their reactions, the stronger the relationship between them. Thus, we weight an edge by the Jaccard Index of the set of reactions of each business, representing an index of affinity or similarity between the two sets.

In a more formal way, consider $ B=\{b_1,b_2,...,b_{n_b}\}$ being the set of all businesses, where $ n_b $ is the total number of businesses in the dataset. Now consider $ U_i $ being the set of all users who reacted to the \textit{i-th} business. Thus, the graph is defined as in (\ref{eq:Graph}):

\begin{equation}\label{eq:Graph}
    BusinessGraph = (V,E,W),
\end{equation}

where vertices are businesses, $ V = B $; edges exist if businesses have a minimum of users’ reactions in common, $ E = \{(i,j) : |U_i \cap U_j|>lowerBound\} $; and the weights of the edges are represented, as in Equation (\ref{eq:GraphWeight}), by the \textit{Jaccard Index}:

\begin{equation}\label{eq:GraphWeight}
W(i,j) = \begin{cases}
            \frac{|U_i \cap U_j|}{|U_i \cup U_j|} & \text{if } (i,j) \in E \\
0 & \text{if } (i,j) \notin E
\end{cases}
\end{equation}

\subsection{Filters and Graph Consistency}

In order to increase the consistency of the information about the graph structure, it is considered two essential filtering steps: a reactions filter, and a weak edges filter.

First, the reaction filter eliminates negative reactions (of the type \textit{Angry} and \textit{Sad}), because for possible partnerships between businesses what matters are positive reactions. Then, the reaction filter eliminates users who do not frequently express themselves about business in $ B $. So users with two or fewer reactions are eliminated from the model, \textcolor{black}{ leaving the filter lower bound as 3 reactions. }

On the other hand, the filter also eliminates users with too many reactions \textcolor{black}{ for two reasons. } (i) The proportion of the number of edges $ a $ that a user with $ m $ reactions generates in the graph is quadratic $ a = \frac{m(m-1)}{2} $\textcolor{black}{ , therefore, for instance, users with $500$ reactions generates $124,750$ edges, unbalancing their influence over users with few reactions. (ii) } Users with too many reactions could be robots (\textit{bots}), a problem that appears in several Web systems \cite{tasse2017state}. \textcolor{black}{ In order to maintain a fair balance between representativeness over potential problems, $99.9\%$ of the reactions (for users with $3$ or more reactions) in the original database are kept. The filter upper bound is calculated counting the proportion of reactions $R_{3-x}$ (reactions from $3$ to $x$) over $R_{3-\infty}$ (all reactions above $3$), as in equation \ref{eq:reacProp}.}

\textcolor{black}{\begin{equation}\label{eq:reacProp}
    \frac{R_{3-x}}{R_{3-\infty}} = 99.9\%
\end{equation}}

\begin{figure}
    \centering
    \includegraphics[width=0.95\textwidth]{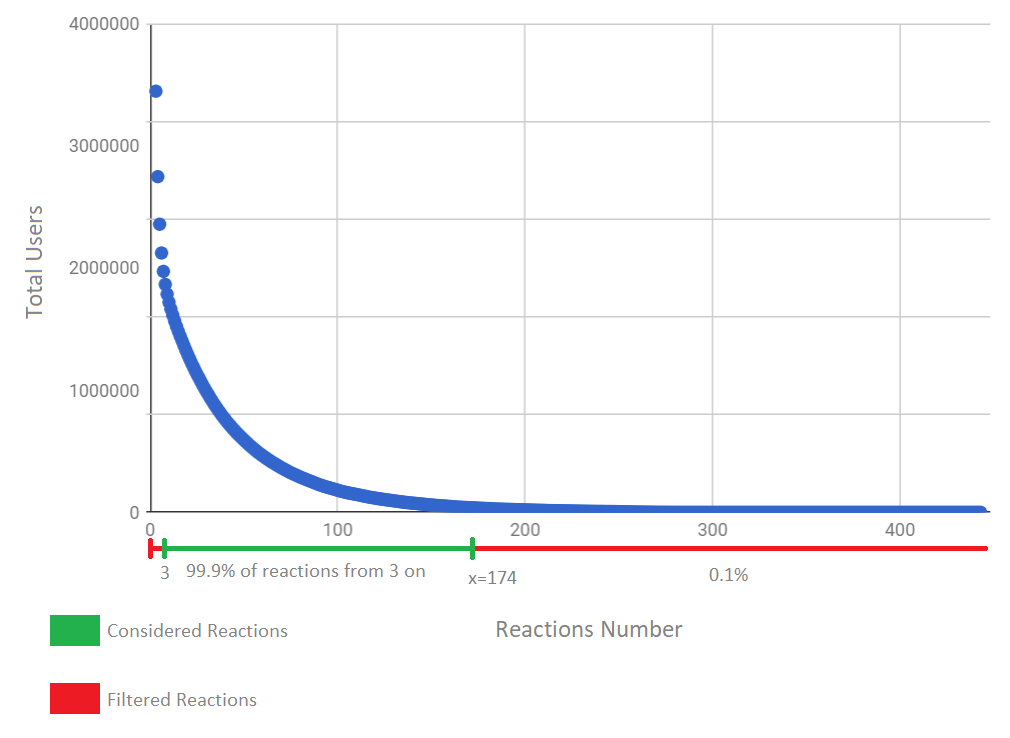}
    \caption{Distribution of the number of users by number of reactions.}
    \label{fig:UserLikeDistribution}
\end{figure}

As depicted in Figure \ref{fig:UserLikeDistribution}, which shows the distribution of total users per number of reactions, \textcolor{black}{ the filter upper bound calculated for the studied dataset is $x=174$; therefore the reaction filter considers only users who reacted from $3$ to $174$ times. } After the reaction filtering process, the resulting dataset has 220 million reactions. 


\textcolor{black}{ Once the reactions are filtered, } the weak edge filter removes possible noises in the graph structure, thus removing edges classified as weak edges. We classify an edge as a weak edge by performing a random experiment, in which the typical reactions are randomly \textcolor{black}{ and uniformly } distributed among all possible edges in the graph, forming a random graph. Then\textcolor{black}{ , after this simulation process, } the edges of the original structure with similar weight to the experiment's graph can be considered weak edges.

\textcolor{black}{ When reactions are uniformly distributed among all possible edges, the weight of any edge in the random graph follows a binomial distribution. } The expected value and the variance of the weight of any edge in the random graph (with binomial distribution) are, respectively, given by Equations (\ref{eq:ExpectedEdge}) and (\ref{eq:VarianceEdge}).

\begin{equation}\label{eq:ExpectedEdge}
    \mu = E[X] = \frac{n_r}{n_c}
\end{equation}
\begin{equation}\label{eq:VarianceEdge}
    \sigma^2 = Var[X] = \frac{n_r}{n_c} \big(1-\frac{1}{n_c}\big)
\end{equation}
Where, $ n_r=\frac{\sum_{i,j:i \neq j} |U_i \cap U_j|}{2} $ \textcolor{black}{ is the sum of weights in the original graph }, $ n_c = \frac{n_b (n_b-1)}{2} $ \textcolor{black}{ is the number of all possible edges } and $ n_b $ is the number of businesses in the dataset.

In this way, an edge between businesses $ i $ and $ j $ is weak if: $ |U_i \cap U_j| \leq lowerBound $, as defined in Equation (\ref{eq:LowerBound}), \textcolor{black}{ following the $3\sigma$ statistics for the binomial distribution, which includes $99.73\%$ of random edges }.

\begin{equation}\label{eq:LowerBound}
    lowerBound = \mu + 3\sigma
\end{equation}

For the data collected in this study the calculated values were: $ \mu = 120.289 $, $ \sigma = 10.96 $. Thus, $ lowerBound = 153.195$. Considering the dataset of this study, $ \num[group-separator={,}]{978410} $ edges were eliminated, resulting in a total of $ \num[group-separator={,}]{223939} $ edges in the graph with less probability of being random noise.

\subsection{Detection of Business Communities}\label{subsec:communitiesdetection}

Given a consistent network of business relationships, an essential step in achieving the study's goal is to detect business communities. As the business graph diameter is $4$, therefore not sparse considering the number of nodes and edges, community detection algorithms based on searching cliques or dense subgraphs with optimal solution, such as the Clique Percolation Method, have a very high computational \textcolor{black}{ time and space } complexity; therefore they are not applicable in this study.

Raghavan et al. \cite{raghavan2007near} proposed a community detection algorithm based on label propagation \textcolor{black}{ (LP) }, which iteratively uses the exchange of labels between adjacent vertices in such a way that promotes convergence of labels. \textcolor{black}{ One significant advantage is that } this algorithm operates in almost linear time, which makes it tractable for dense graphs. \textcolor{black}{ Also, since previous information about the graph and communities are not available, another advantage is that this LP-based algorithm does not need previous information, such as heuristics of communities present in the input network, required in other methods cited in \cite{raghavan2007near}. It is also true that LP-based algorithms sometimes give different solutions due to the random steps performed. To enhance the stability of the final solution this LP-based algorithm has aggregation steps to combine different solutions. }

For the problem addressed here, it is interesting that communities have at least four businesses (\textit{minSize}), and a maximum of thirty businesses (\textit{maxSize}), since huge communities lose cohesion in possible recommendations. Therefore, an iterative algorithm, Algorithm \ref{alg:businessCommunity}, is proposed for the detection of communities of businesses. The entries of this algorithm are the business graph (\textit{BusinessGraph}), the minimum size (\textit{minSize}) and maximum size (\textit{maxSize}) of the communities, and the output is a set of business communities. The algorithm performs the following key steps:

\begin{itemize}
    \item Detection of communities with the algorithm described in \cite{raghavan2007near};
\item Business communities with size of \textit{minSize} and \textit{maxSize} are saved for final return;
\item The union of detected communities, and not saved for return, composes a new graph, named \textit{G};
\item From this \textcolor{black}{ new } graph \textit{G}, weak edges are cut forming the graph for the next iteration.
\end{itemize}

\begin{algorithm}[tbh]
\label{alg:businessCommunity}
\caption{Business Communities Detection Algorithm.}
    \Indm
    \KwData{BusinessGraph,minSize,maxSize}
    \KwResult{Set of Communities of BusinessGraph}
    \Indp
    $G \leftarrow BusinessGraph$ \\
    $allCommunities \leftarrow \emptyset$\\
    $minEdge \leftarrow \min_{i,j \in G} W(i,j)$\\
    $counter \leftarrow 1$\\
    \While{$|G|>minSize$}{
    $counter \leftarrow counter + 1$\\
    \textcolor{gray!65}{/* Method of \textit{\cite{raghavan2007near}} */}\\
    $detectedComm \leftarrow labelPropCommDetection(G)$\\
    $G \leftarrow empty\ graph$\\
    \For{$c \in detectedComm$}{
         \eIf{$|c|>minSize$ and $|c|<maxSize$}{
         $allCommunities \leftarrow allCommunities \cup \{c\}$
     }{
         $ G \leftarrow G \cup c$
     }
     }
         remove all edges of $G$ with $W(i,j)<minEdge * counter$\\
    \If{no edges removed from G}{
    $break$
}
    }
\Return $allCommunities$;
\end{algorithm}

The communities detected according to Algorithm \ref{alg:businessCommunity} are subgraphs that tend to be dense (many edges), so businesses within the same community have a greater cohesion than businesses randomly chosen in the graph. This cohesion is formed by spontaneous user reactions, without additional information that could include a bias in communities detected. \textcolor{black}{ The time complexity of Algorithm \ref{alg:businessCommunity} is $O(|V|+|E|)$, where $G=(V,E)$. }

\subsection{Clustering of Business Communities} \label{sec:clusteringBusCom}


In this section, it is described the steps to cluster business communities by their similarity. As we are dealing with Facebook data, the clustering process is done considering categories of businesses given by Facebook. However similar processes could be done with data from any other social media platform.

Facebook classifies all businesses into one of seven categories: \textit{Interest; Community Organization; Media; Public Figure; Businesses; Non-Business Places}; and \textit{Other}. All of them have subcategories, but the larger number of subcategories ($ 22 $), as well as the greater diversity, are subordinated to the \textit{Business} category. Therefore, we considered all subcategories within \textit{Interest, Community Organization, Media, Public Figure, Non-Business Places,} and \textit{Other} as their parent categories. For instance, all subcategories of \textit{Interest} are transformed into \textit{Interest}.

For the \textit{Business} category, all subcategories have been considered. Under each of them, there are sub-subcategories. The sub-subcategories of \textit{Business} were disregarded, as this level of specialization was not considered interesting for the analysis carried out. The \textit{Advertising or Marketing} subcategory of \textit{Business} has, for example, the sub-subcategories \textit{Advertising Agency} and \textit{Copywriting Service}. In this case, all sub-subcategories within \textit{Advertising or Marketing} are considered \textit{Advertising or Marketing}. We performed the same procedure for all other subcategories within \textit{Business}.

\begin{figure}
    \begin{minipage}{.30\linewidth}
        \centering
        \includegraphics[width=\textwidth]{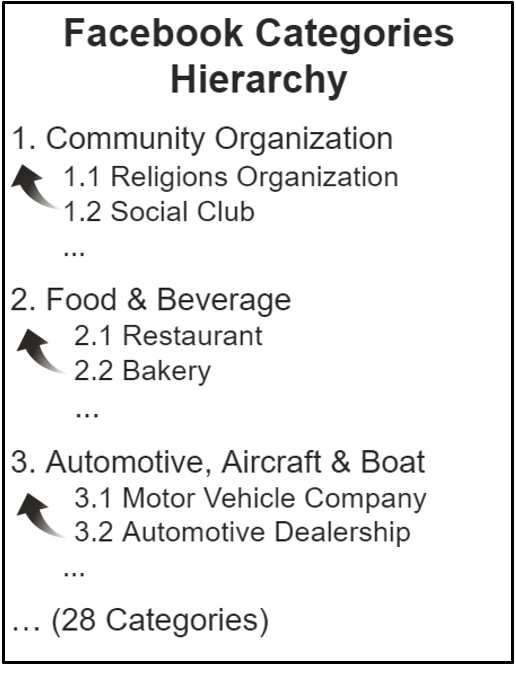}
        \caption{Facebook subcategories renaming process.}
        \label{fig:FaceCategories}
    \end{minipage}%
    \begin{minipage}{.65\linewidth}
        \centering
        \includegraphics[width=\textwidth]{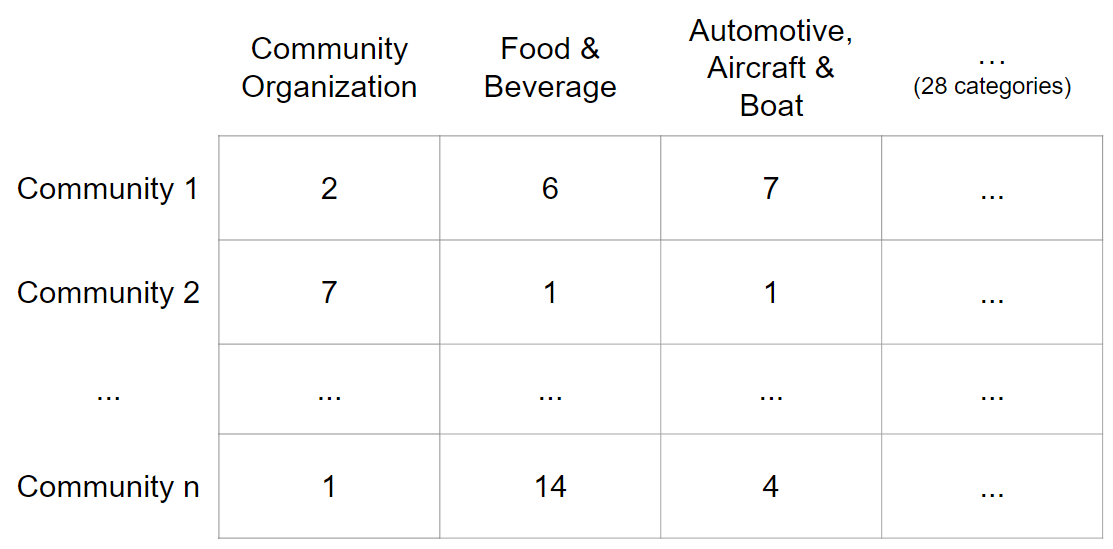}
        \caption{Examples of non-normalized feature vector for each community.}
        \label{fig:CatVectors}
    \end{minipage}
\end{figure}

After doing this process, a total of $ 28 $ business’ category names ($ 6 $ from the parent categories, and $ 22 $ subcategories from \textit{Business}) were obtained, as illustrated in Figure \ref{fig:FaceCategories}. We then built a feature vector with these $ 28 $ categories and counted for each community the number of occurrences of businesses in each of the $ 28 $ categories, as in Figure \ref{fig:CatVectors}. These values are then normalized based on the maximum number of locations in a given feature for each community. With this feature vector (represented formally as $ vector(c) $ for a given community $ c $), it is possible to identify the similarity of different communities and perform a clustering process. The clustering algorithm used was the k-means, with the Euclidean distance \cite{hartigan1975clustering}. For choosing the right $ k $ parameter of k-means algorithm, several different values were tested, and the value with the smallest sum of squared errors was chosen as the best fit for the data. For the dataset presented in this paper, the best fit was $ k = 8 $; thus this value is kept for this study.

Formally, k-means receives a set of all communities (called $allCommunities $ as in Algorithm \ref{alg:businessCommunity}) and returns a set of clusters, where clusters are disjoint non-empty subsets of $ allCommunities $: $$ clusters =  kmeans(allCommunities), $$ \centerline{$\forall{cl_1, cl_2 \in clusters ; cl_1 \neq cl_2}, \quad cl_1 \cap cl_2 = \emptyset, \quad cl_1 \neq \emptyset \quad \textrm{and} \quad cl_2 \neq \emptyset.$}

\subsection{Business Outlier Detection} \label{sec:outlierdetec}

To detect possible businesses outliers inside business communities, first a clustering of business communities, as described in Section \ref{sec:clusteringBusCom}, has to be performed. In possession of clusters of business communities, the detection is based on \textit{cluster centroids}, which represent the common proportion of business categories for each cluster, and are a central piece in the business outlier detection. In a high level of abstraction, if communities differ significantly from its cluster centroids, they have some outlier businesses inside them. Formally, the cluster centroid is the average vector ($ \in {\rm I\!R^{28}} $) among all the communities from that cluster, as in Equation \ref{eq:centroid}:

\begin{equation}\label{eq:centroid}
    \forall cl \in clusters, \quad centroid(cl) = \frac{\sum_{c\in cl}{vector(c)}}{|cl|}
\end{equation}


Cluster centroids are used to build \textit{cluster signatures}. A cluster signature is a set of the most representative business categories for a particular cluster. Each category (dimension) inside the cluster centroid vector represents a certain percentage of all categories present in the vector so that all percentages sum up to 100\%. Therefore, picking the categories with the greatest percentages so that their percentages sum up to a threshold ($>50\%$), makes them the set of the most representative categories for that cluster. For example, a cluster could have as its signature the categories \textit{Food and Beverage}, \textit{Shopping and Retail} and \textit{Entertainment}, because they meet a 70\% threshold defined in a particular application.



More formally, Algorithm \ref{alg:getSignature} specifies how to capture a \textit{cluster signature} given two inputs: A vector $v \in {\rm I\!R^n}$, representing the centroid, where $n$ is the total number of categories; and a $threshold \in \quad ]0.5,1] $, a number representing the minimum percentage considered as majority. As a result, Algorithm \ref{alg:getSignature} returns a set containing the greatest dimensions, i.e. categories, of the vector $v$ that corresponds to the closest possible number to $threshold*100\%$ of all dimensions. \textcolor{black}{ Its time complexity is $O(|v|^2)$. }


\begin{algorithm}[tbh]
    \label{alg:getSignature}
    \caption{Function that returns the signature (greatest dimensions) of the vector.}
    \SetKwFunction{FGSig}{getSignature}
    \SetKwProg{Fn}{Function}{:}{}
    \Fn{\FGSig{$v$, $threshold$}}{
        $s \leftarrow \sum_{i} v_i$\\
        $accThrs \leftarrow 0$\\
        $signature \leftarrow \emptyset$\\
        \For{$i \in \{1,2,...,|v|\}$}{
            $m \leftarrow max(v)$            \textcolor{gray!65}{/*Max value in vector*/}\\
            $j \leftarrow argmax(v)$          \textcolor{gray!65}{/* Category (index) of the vector's maximum value*/}\\
            \eIf{$|\frac{m+accThrs}{s} - threshold| < |\frac{accThrs}{s} - threshold| $}{
                $signature \leftarrow signature \cup j$\\
                $accThrs \leftarrow accThrs + m$\\
                $v_j \leftarrow 0$ \textcolor{gray!65}{/*In next iteration, max(v) is the next greatest*/}\\
            }
            {
                \textbf{break}
            }
        }
    }
    \KwRet $signature$\\
\end{algorithm}


Finally, Algorithm \ref{alg:outlier} is responsible for tagging all businesses which categories are not included in its corresponding cluster signature, therefore considered as outliers. This algorithm takes as input a set of clusters and returns the same input structure, however with additional outlier tags. Note that firstly it calls Algorithm \ref{alg:getSignature}, to get each cluster signature, and next it tags businesses which categories are not in its cluster signature. Note that each community has a vector (as in Figure \ref{fig:CatVectors}), and each cluster has a centroid (as in Equation \ref{eq:centroid}), both are captured in Algorithm \ref{alg:outlier} as $vector(community)$ and $centroid(cl)$, respectively. \textcolor{black}{ In order to capture the most representative categories of each cluster, the function $getSignature$ (Algorithm \ref{alg:getSignature}) is called. For this function, if the $threshold$ value is high (e.g., $0.9$), then too many categories are considered in the cluster signature, on the other hand, if the value is low (e.g., $0.5$) then there is a chance that no category is considered  in  the  cluster  signature. According to our empirical analysis, the value  $0.7$  represents a good balance; therefore, we consider it as the threshold. }

\begin{algorithm}[tbh]
\label{alg:outlier}
    \Indm
    \KwData{Clusters - set containing business community clusters}
    \KwResult{Clusters with tagged businesses}
    \Indp
    \textcolor{black}{ $taggedClusters \leftarrow \emptyset$ }\\
    \For{$cl \in clusters$}{
        \textcolor{black}{ $newCluster \leftarrow \emptyset$ }\\
        $clSignature \leftarrow getSignature(centroid(cl),0.7)$\\
        \For{$community \in cl$}{
            $vc \leftarrow vector(community)$ \\
            \For{$i \in \{1,2,..,|vc|\}$}{
                \If{
                    $vc_i>0$ and $i \not\in clSignature$
                }{
                    \textcolor{gray!65}{/*$tagBusiness(community,cat)$ tags businesses of category $cat$ inside $community$*/}
                    \textcolor{black}{ $newCluster \leftarrow newCluster \cup \{tagBusinesses(community,i)\}$ }  \\
                }
            }
            \textcolor{black}{ $taggedClusters \leftarrow taggedClusters \cup \{newCluster$ }\}
        }
    }
\Return \textcolor{black}{ $taggedClusters$ }
\caption{Outlier Detection Algorithm. }
\end{algorithm}

\textcolor{black}{ The time complexity of Algorithm \ref{alg:outlier} is $O(|clusters|*|cat|^2 + |allCommunities|*|cat| + |V|)$, where $G=(V,E)$, $cat$ is the set of all categories, and $allCommunities$ represent the set of all communities being studied. }

\section{Results \textcolor{black}{ and Discussions }}
\label{sec:results}

The two outputs of the framework are the tagged communities, i.e., tagged as an outlier or not, that includes the target business (informed by the user), and the \textit{egonet} of the target business, consisting of a subgraph of all edges of the target business (see Figure \ref{fig:Fluxograma}). As there are considerably large \textit{egonets} for individual businesses, the \textit{egonet} size in this study was limited to a maximum of seven adjacent vertices (with the strongest edges) plus the target business. Note that this parameter could be adjusted for each case under study.

\begin{figure}[t]
    \centering
    \includegraphics[width=0.85\textwidth]{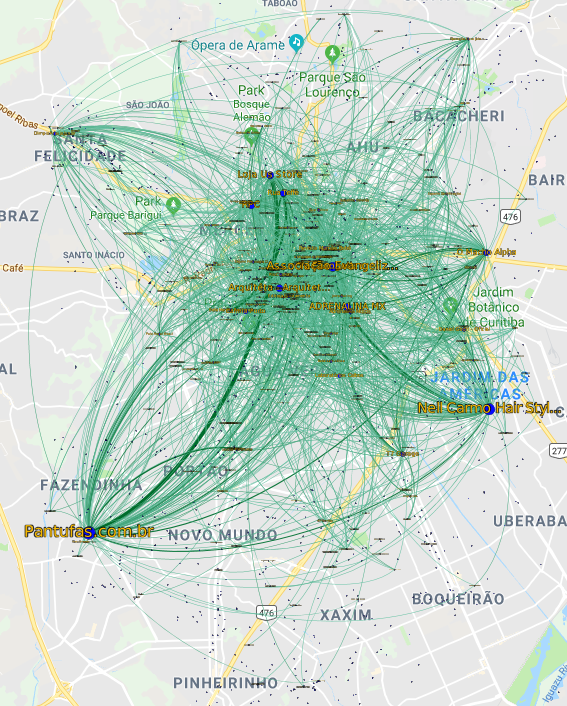}
    \caption{Partial view of the business graph on the map of Curitiba.}
    \label{fig:FaceGraphMap}
\end{figure}

Figure \ref{fig:FaceGraphMap} illustrates the complete graph, constructed by the proposed framework, shown on a map of the city of Curitiba. To ease the visualization, we only show edges with more than $ 1,500 $ common reactions. The thicker the edge, the stronger the relationship between nodes. We show each node in the graph according to the location of the business it represents. Note that there are vertices far from the city center with a considerably high density of edges going towards the city center, indicating a strong activity also in outlying neighborhoods.

An important detail was captured during analysis. The nodes ``Prefeitura de Curitiba'' (N1)  and ``RPC'' (N2) have a strong influence on business graph (as shown in Appendix \ref{app:reacDatabase} by Tables \ref{tab:graphNodeRanking} and \ref{tab:graphEdgeRanking}). N1 represents the city hall of Curitiba, and it is a very popular Facebook page \footnote{https://www.facebook.com/PrefsCuritiba.} not only in the city of Curitiba but also nationally. N2 is the largest TV channel in Curitiba. Both strongly impact the analysis carried out by this study. N1 does not represent a business and N2 is an obvious TV media partnership. Their edges influence on the business relationship graph potentially hides interesting smaller business relationships. In order to favor more valuable relationships,  the analysis was performed without both nodes.

\begin{figure}[t]
    \centering
    \subfigure[Community of businesses related to leisure. (No businesses tagged as outliers)]{
        \label{fig:LeisureCommunity}
        \includegraphics[width=0.45\textwidth]{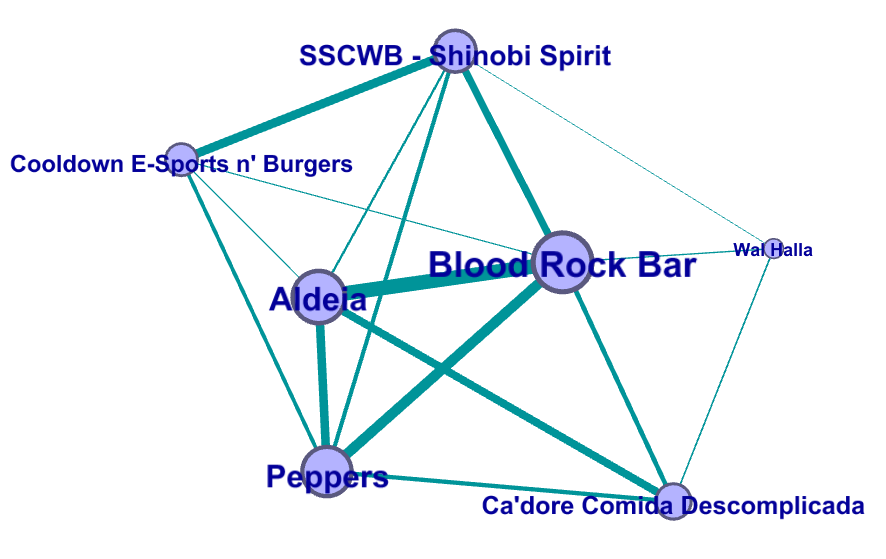}
    }
    \subfigure[Community of businesses related to fashion. (No businesses tagged as outliers)]{
        \label{fig:FashionCommunity}
        \includegraphics[width=0.45\textwidth]{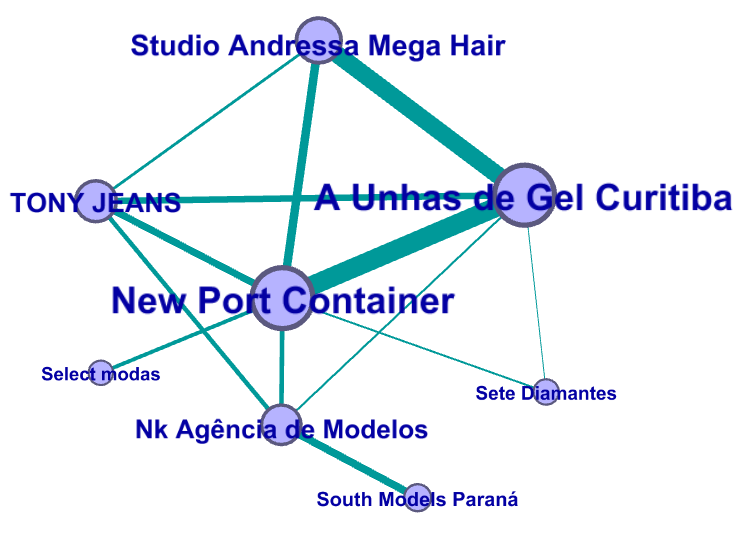}
    }
    \subfigure[Another Community of businesses related to Fashion. The business tagged in red is a health plan related business and it was considered an outlier by Algorithm \ref{alg:outlier}]{
        \label{fig:FashionCommunity2}
        \includegraphics[width=0.75\textwidth]{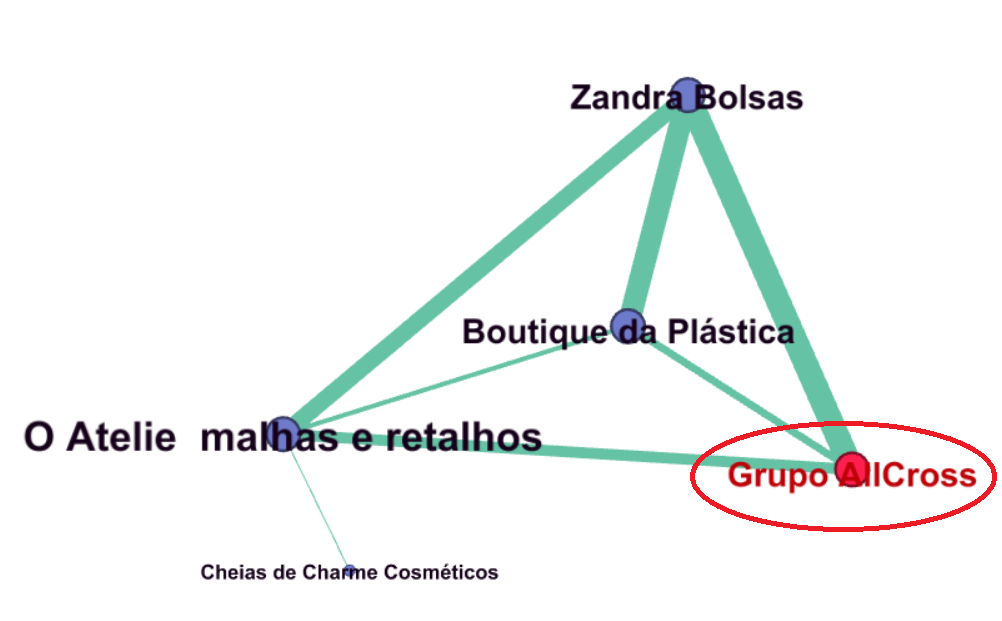}
    }
    \caption{Communities detected by Algorithm \ref{alg:businessCommunity}}
\end{figure}

As this graph is quite large, the extraction of useful information becomes complicated to human eyes, justifying the extraction of communities and \textit{egonets}. After running Algorithm \ref{alg:businessCommunity} with the parameters considered, $ 144 $ communities were detected, each ranging from $ 4 $ to $ 30 $ businesses located in the city of Curitiba. Figure \ref{fig:LeisureCommunity} illustrates a community containing entertainment businesses (e.g., ``Blood Rock Bar'', and ``SSCWB - Shinobi Spirit'') and food businesses (e.g., ``Ca'dore Comida Descomplicada''), so they are businesses united by the ``leisure'' context. The two communities illustrated in Figures \ref{fig:FashionCommunity} and \ref{fig:FashionCommunity2} both have businesses bound together by the context that can be called ``fashion'', because it contains businesses from the beauty salon sector (e.g., ``Studio Andressa Mega Hair'' and ``Cheias de Charme Costméticos''), modeling agencies (e.g., ``Nk Agencia de Modelos'' and ``South Models Parana'') and fashion stores (e.g., ``TONY JEANS'' and ``Zandra Bolsas'' ). We can note that, even though both the business network construction (see Section \ref{sec:graph}) and the Algorithm \ref{alg:businessCommunity} did not use any information of the businesses themselves, all communities detected have similar strong semantics that binds businesses together inside each community.

\begin{figure}[t]
    \centering
    \subfigure[Cluster 1. \textcolor{black}{ Arts \& Entertainment; Local Service; Media News Company }]
    {\includegraphics[width=.30\textwidth]{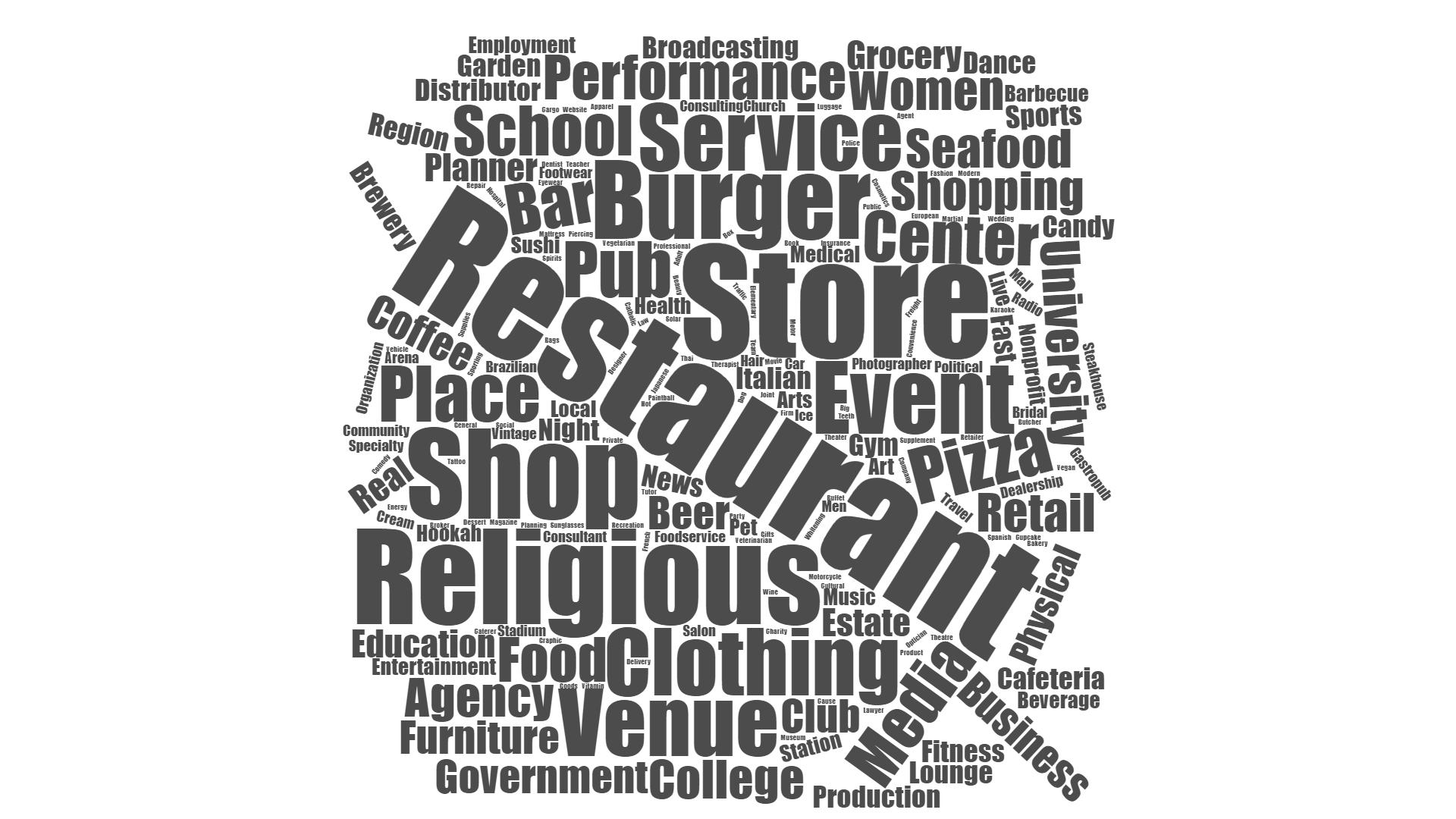}} 
    \subfigure[Cluster 2. \textcolor{black}{ Beauty Cosmetic \& Personal Care; Shopping \& Retail; Other. }]
    {\includegraphics[width=.32\textwidth]{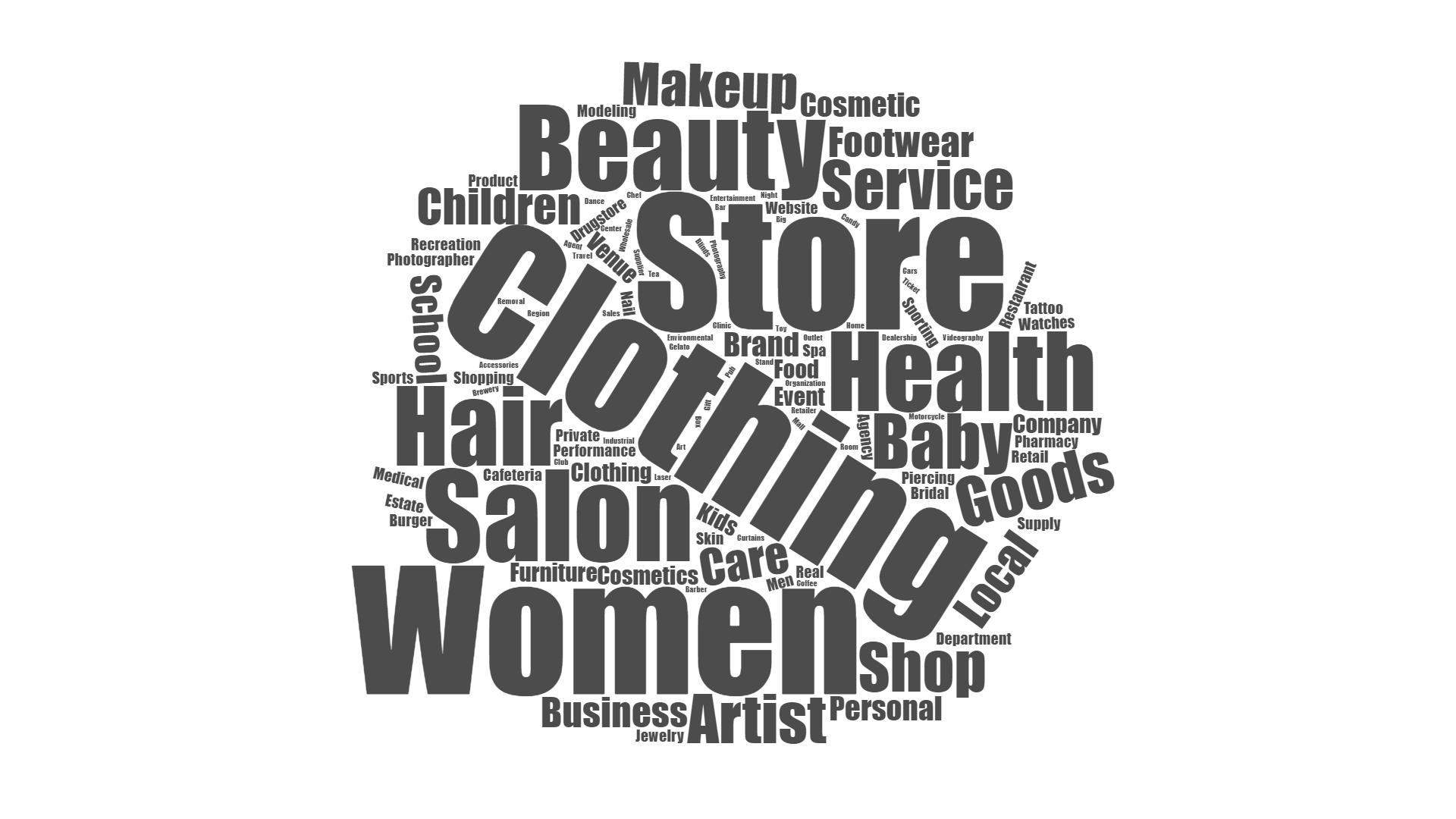}}
    \subfigure[Cluster 3. \textcolor{black}{ Automotive, Aircraft \& Boat; Commercial \& Industrial; Non-profit Organization. }]
    {\includegraphics[width=.30\textwidth]{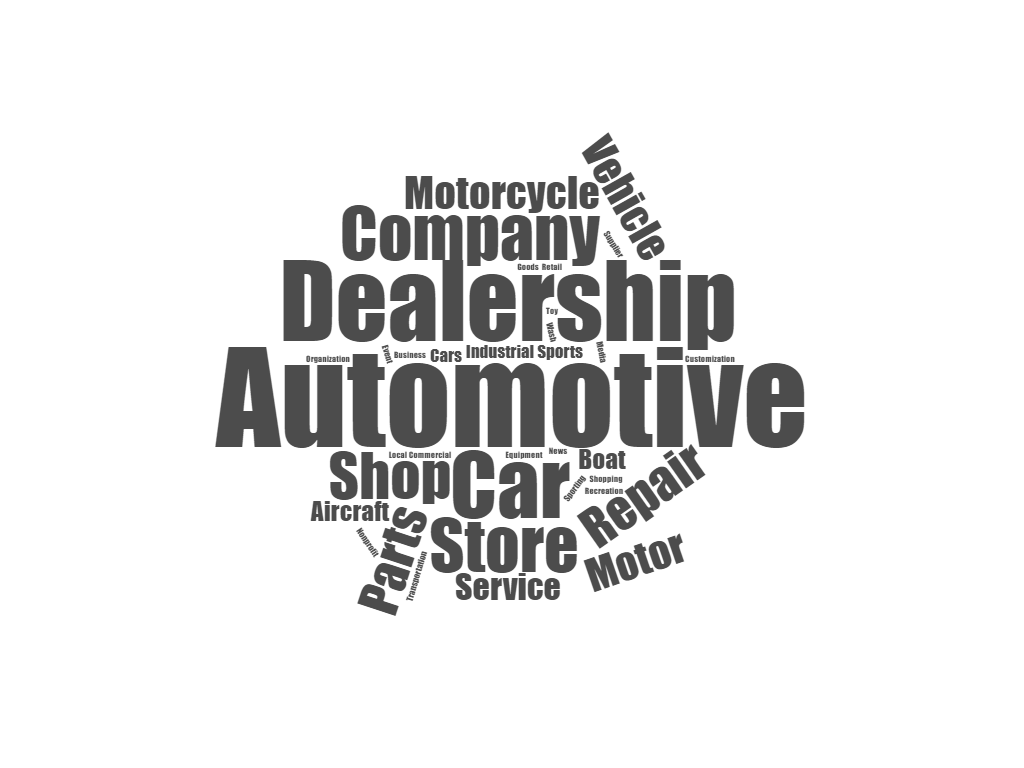}}
    \subfigure[Cluster 4.\textcolor{black}{  Media; Hotel \& Lodging; Other }]
    {\includegraphics[width=.30\textwidth]{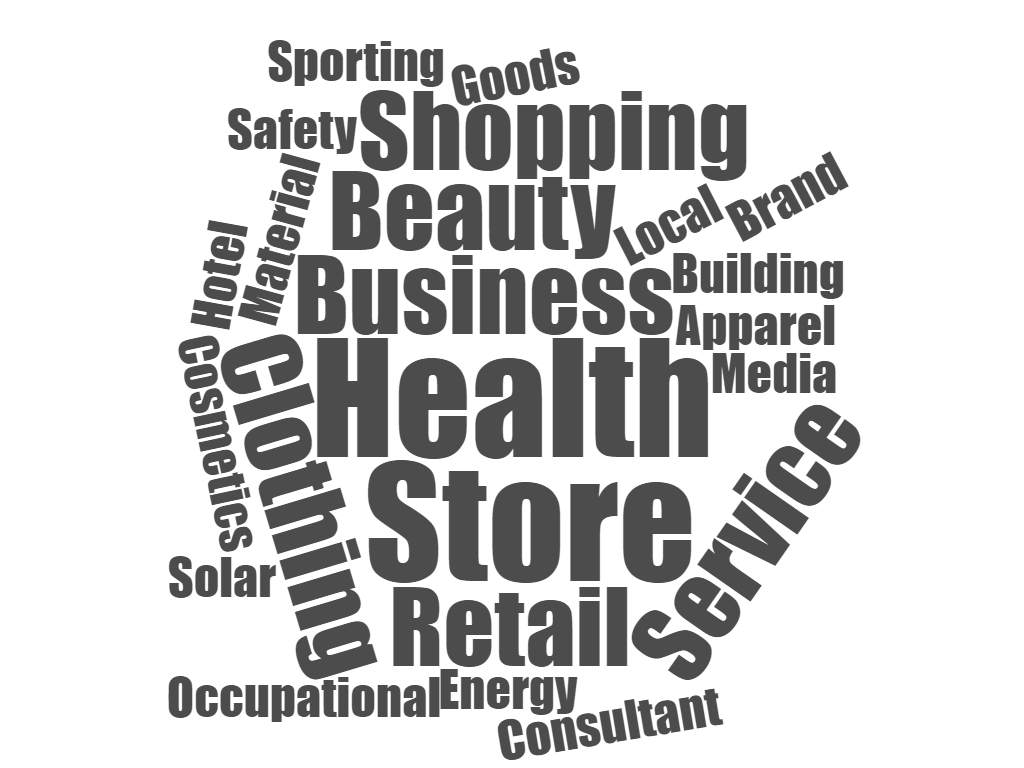}}
    \subfigure[Cluster 5. \textcolor{black}{ Education; Public Figure; Other. }]
    {\includegraphics[width=.30\textwidth]{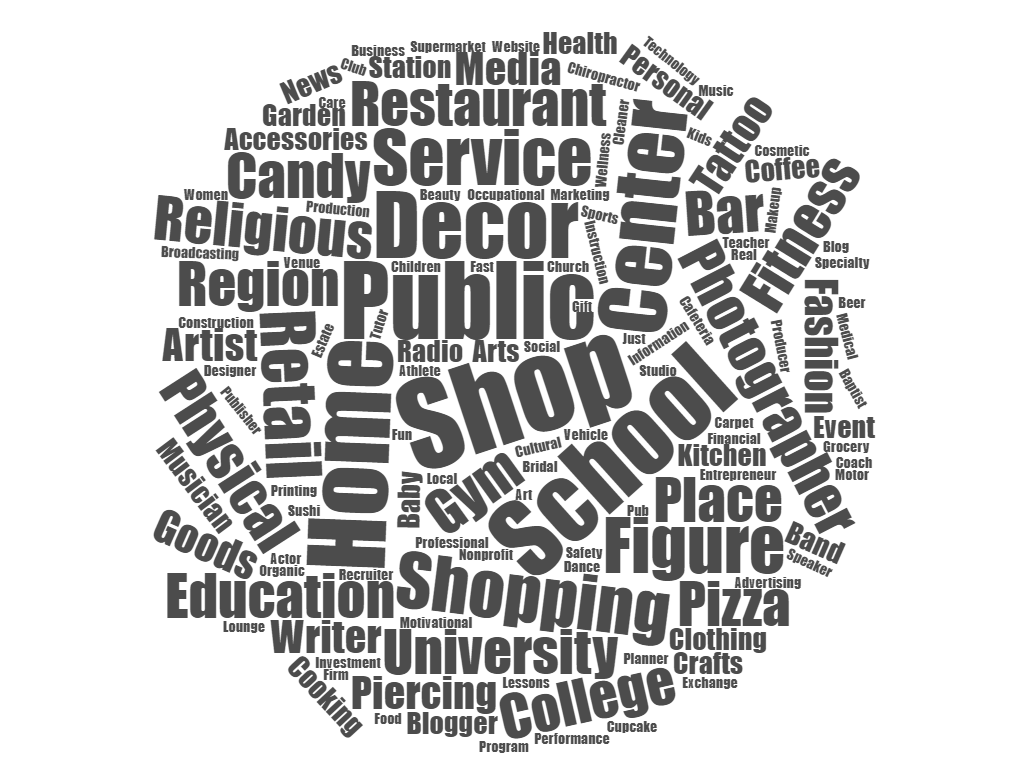}}
    \subfigure[Cluster 6. \textcolor{black}{ Sports \& Recreation; Travel \& Transportation; Advertising Marketing. }]
    {\includegraphics[width=.30\textwidth]{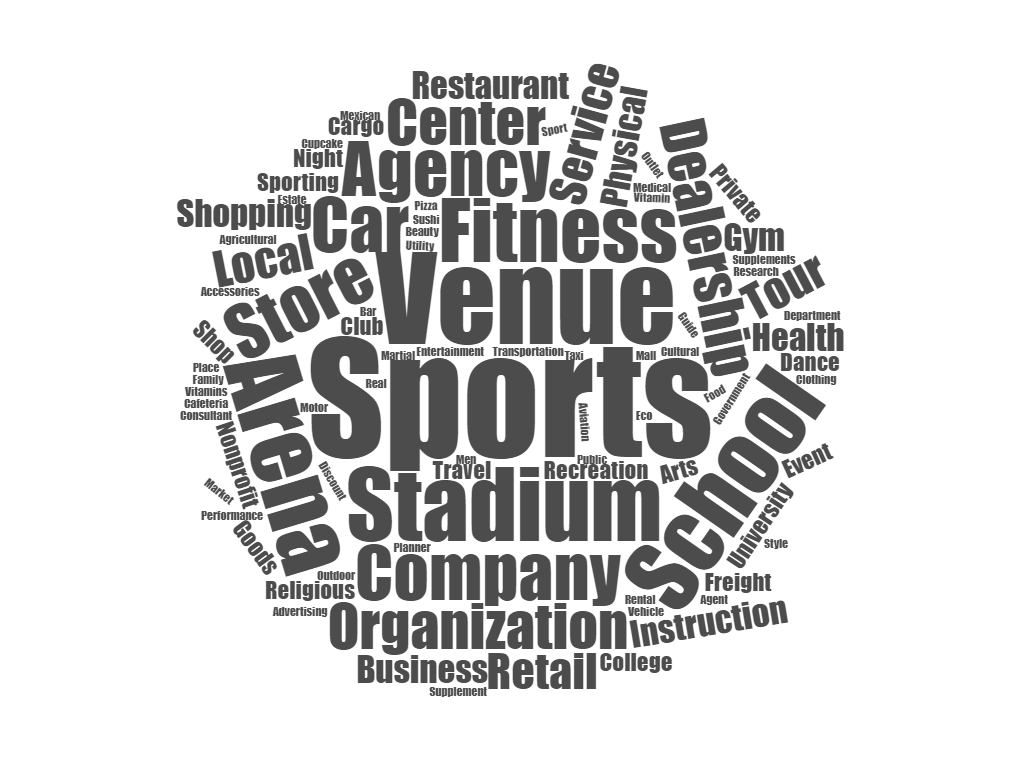}}
    \subfigure[Cluster 7. \textcolor{black}{ Science, Technology \& Engineering; Hotel \& Lodging; Advertising Marketing. }]
    {\includegraphics[width=.30\textwidth]{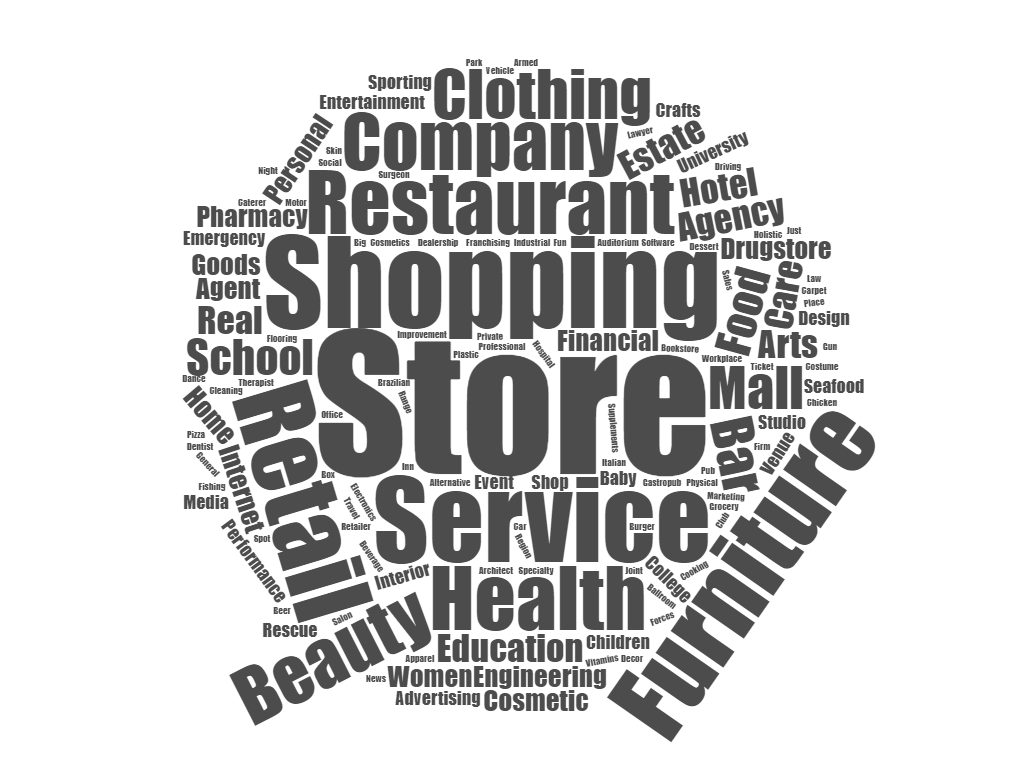}}
    \subfigure[Cluster 8. \textcolor{black}{ Local Service; Shopping \& Retail; Medical \& Health. }]
    {\includegraphics[width=.40\textwidth]{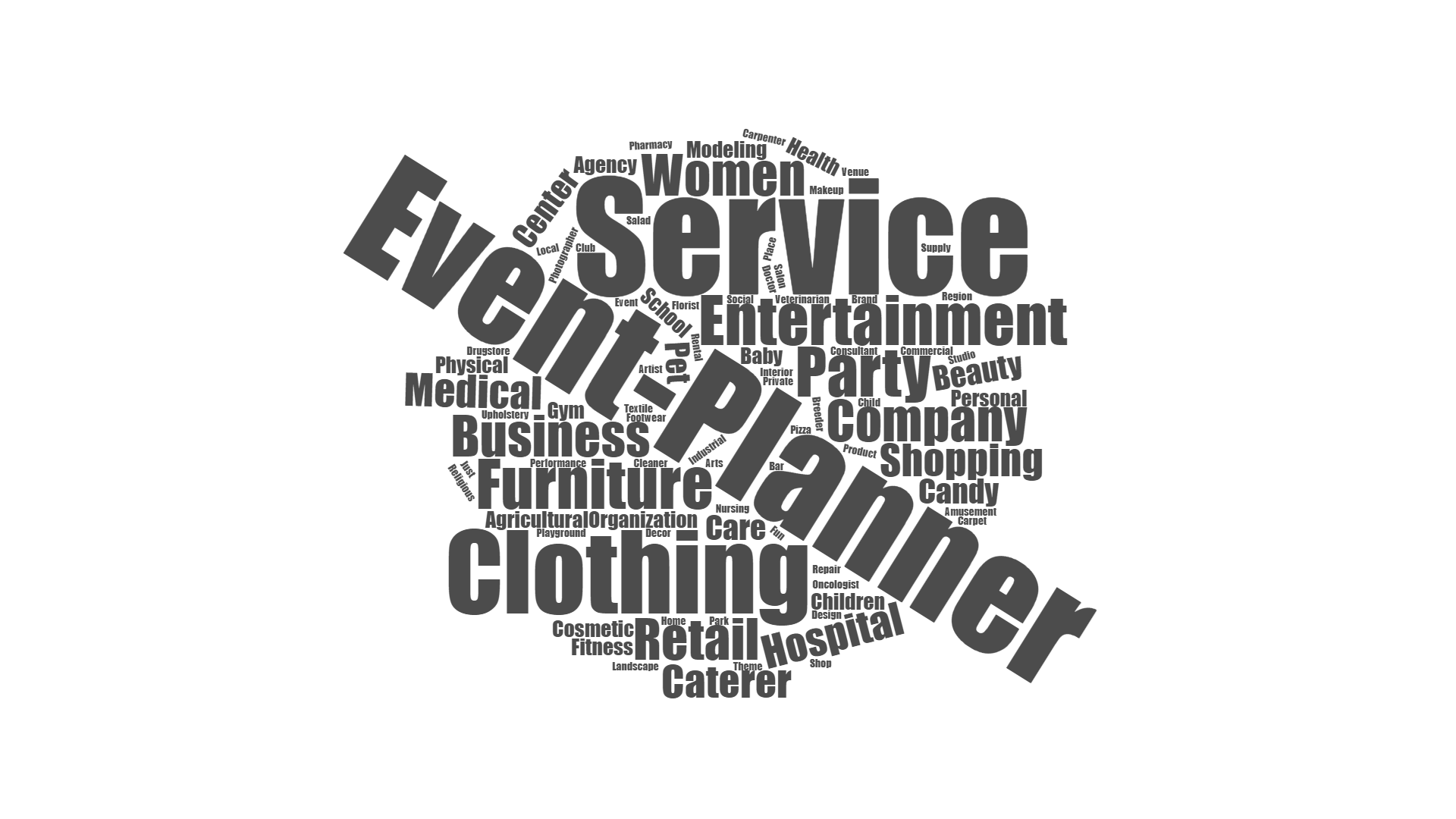}}
    \caption{Word clouds for categories of similar community clusters. \textcolor{black}{ Each cluster legend contains the three biggest super categories of its signature according to Algorithm \ref{alg:getSignature}. }}
    \label{fig:wordcloudsGrupos}
\end{figure}

\begin{figure}[ht]
    \includegraphics[width=\textwidth]{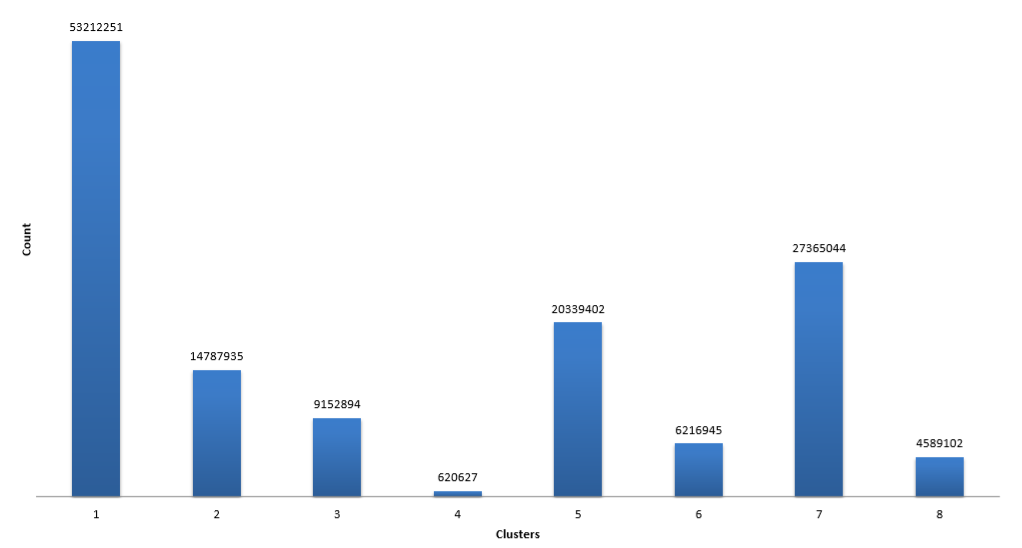}
    \caption{A chart presenting user reaction number per cluster}
    \label{fig:clusterLike}
\end{figure}

The business category clustering analysis can illustrate those contexts in a more general view, considering all communities detected. The clustering step, then, unites all similar communities, by business categories, in eight different clusters (for $ k = 8 $ as discussed in Section \ref{sec:clusteringBusCom}). To illustrate the clusters, eight word clouds of businesses' categories, inside each cluster, were generated and shown in Figure \ref{fig:wordcloudsGrupos}. Besides, Figure \ref{fig:clusterLike} shows the number of user reactions in each cluster.

In those word clouds, we did not perform the subcategory renaming process made to execute K-means, so we considered the original names. Note the surprising similarity between the categories in each group. For example, Cluster 1 is related to leisure, containing predominantly food, drink and entertainment businesses, Cluster 2 contains most businesses related to beauty and style, while Cluster 3 is more related to establishments about automotive products and services. This analysis shows the existence of a predominant context in each community. Taking into account information in Figure \ref{fig:wordcloudsGrupos} and Figure \ref{fig:clusterLike}, it is possible to notice that the most popular contexts are leisure and food, represented by Cluster 1, followed by shopping malls, represented by Cluster 7.


Knowing that there is a tendency of having a predominant context of business in communities, outliers, i.e., business outside the predominant type of business, can be useful for decision makers. Next, we present results in this direction following the procedures described in Section \ref{sec:outlierdetec}. 

The community illustrated in Figure \ref{fig:FashionCommunity2}, which is a fashion related community (its predominant context), has one outlier inside it, which is the business called ``Grupo AllCross'' (tagged in red). This business is a health plan consultant business, being not part of the ``fashion'' context and, thus, correctly identified as an outlier by Algorithm \ref{alg:outlier}. \textcolor{black}{ As an improvement of the results in \cite{diegoCourb18}, } outliers cannot be ignored in the results presented here, as they might represent non-trivial potential business partnerships. Although outliers are not part of the dominant context, they still have strong connections to businesses from that context.

\begin{figure}[t]
    \centering
    \subfigure[Egonet for the company called Rubiane.]{
        \includegraphics[width=.85\textwidth]{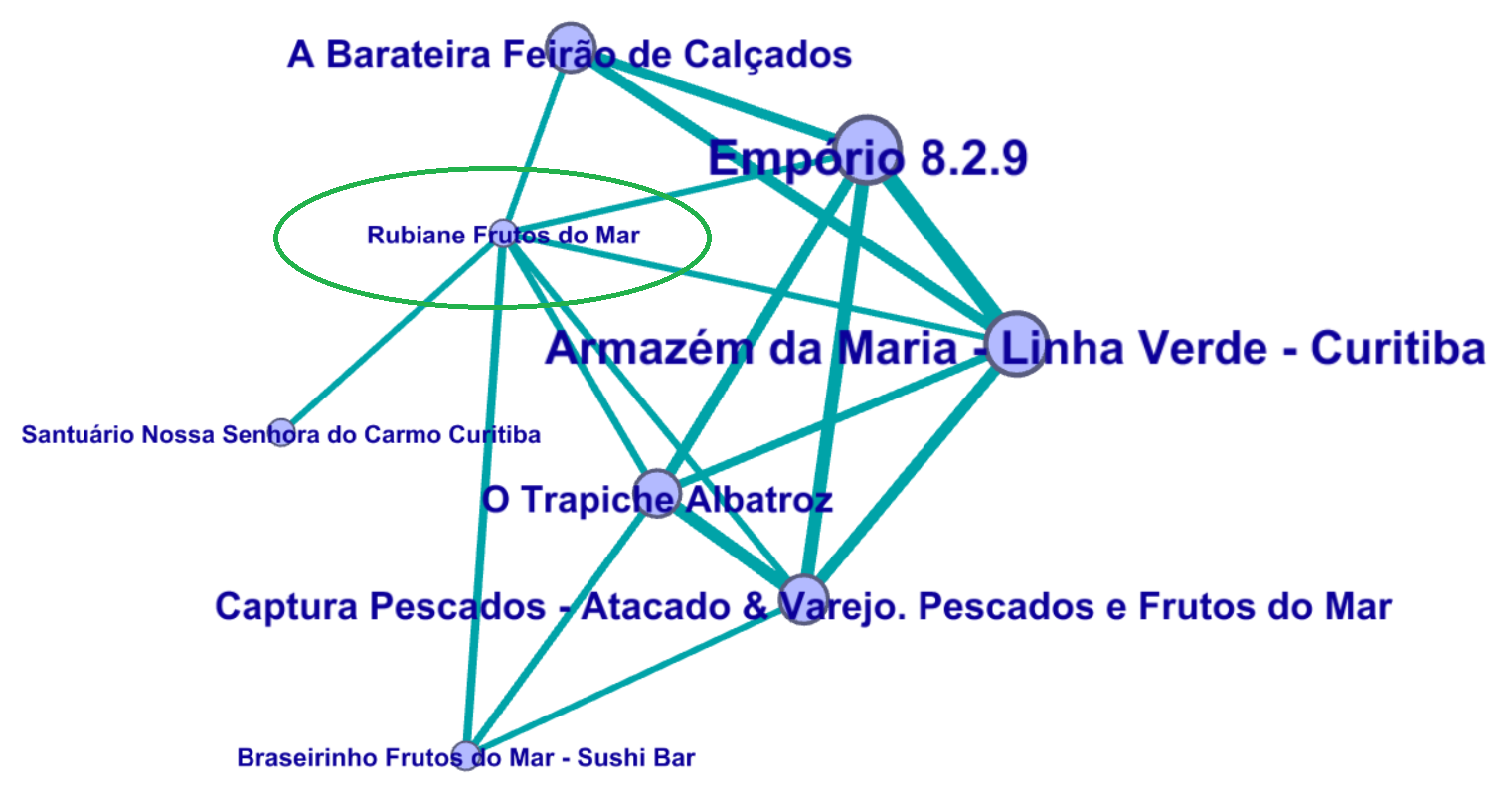}
        \label{fig:EgonetRubiane}
    }
    \subfigure[Community of businesses related to Food (including Rubiane). In red, a tourism related business was tagged as outlier by Algorithm \ref{alg:outlier}.]{
        \includegraphics[width=.85\textwidth]{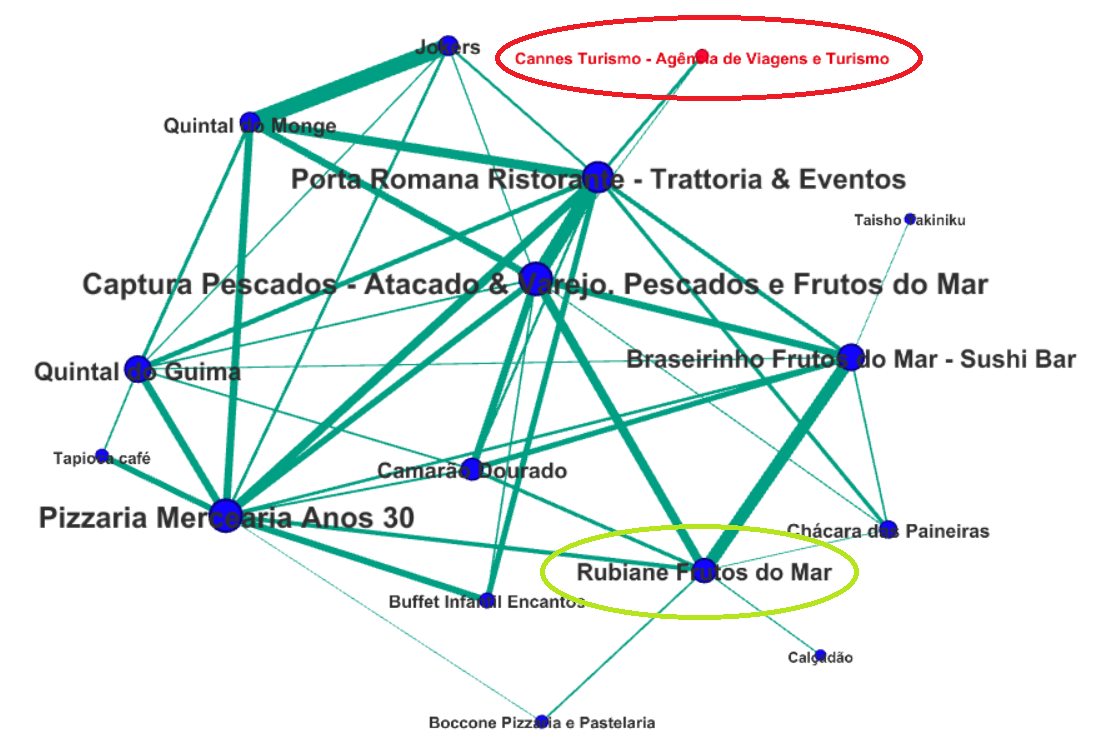}
        \label{fig:FoodCommunity}
    }
    \caption{Framework output for Rubiane.}
\end{figure}

Figure \ref{fig:EgonetRubiane} shows the \textit{egonet} of Rubiane, a seafood restaurant, which was arbitrarily chosen for analysis, and Figure \ref{fig:FoodCommunity} shows a detected community in which Rubiane is included. On the one hand, having the business' \textit{egonet}, it is possible to visualize the direct connections that the target business possesses with other businesses. On the other hand, having communities, it is possible to notice connections that may not be direct to the target business. Since these non-direct connections are within a community (detected by the Algorithm \ref{alg:businessCommunity}), they are cohesive (a dense subgraph) and may represent possible non-trivial partnerships for the business under evaluation. For example, the company ``Quintal do Monge'' does not appear in the Rubiane's \textit{egonet} shown in Figure \ref{fig:EgonetRubiane}, but it appears in a community where Rubiane is also included, shown in Figure \ref{fig:FoodCommunity}. Also in Figure \ref{fig:FoodCommunity} notice that the business called ``Cannes Turismo'' is a tourism related business and was tagged as an outlier by Algorithm \ref{alg:outlier}. 

Rubiane, for instance, could make use of this result to increase its sales by creating business partnerships, such as selling products and services along with the businesses found in the results, as well as marketing partnerships and joint marketing campaigns. For the case of the restaurant analyzed here, we observe that competitors appeared in the same community, for example, ``Braseirinho Frutos do Mar''. For the case involving restaurants, this could be explained by the fact that users tend to attend several restaurants and some may be of the same type. However, this is not a problem with the proposed approach, since the entrepreneur is who decides the best strategy of how to explore the results. Note that a partnership could be made with competing establishments. However, these cases deserve special attention.

\section{Conclusion and Future Work}
\label{sec:conclusion}

The approach introduced in this study aims to provide a new way of identifying significant and non-trivial relations between business, which could ease the laborious task of strategic business partnerships recommendation. This study shows, using large-scale data from Facebook, that the proposed approach could be an important building block for the development of new applications and services, including a business partnership recommender. \textcolor{black}{ Furthermore, the presented results and discussions show that the data available in social media and other platforms are indeed helpful for us to understand the dynamics of the world around us.}

\textcolor{black}{Considering this study as a basis for further research, there are many directions to follow.} For instance, we observe the existence of competing businesses in the same community. As one of the possible implications of this study is to contribute to identifying new business partnerships, it may be interesting to determine a way to detect whether two businesses are competitors to improve the performance of possible recommendations. \textcolor{black}{ Besides, it is interesting to evaluate the results for a new dataset, especially in one representing a different culture. } In addition, it is essential to perform a qualitative evaluation considering business owners or decision makers of the studied businesses. This is essential to understand how to explore the results in practice better. Another direction is to consider the temporality of the reactions, to evaluate, for example, the temporal correlation in the communities. 

\begin{backmatter}

\section*{Competing interests}
  The authors declare that they have no competing interests.

\section*{Author's contributions}
    D.P.T. and T.H.S. designed the model and the computational framework and analyzed the data. D.P.T. carried out the implementation. D.P.T., A.T.F., and T.H.S. wrote the manuscript with input from all authors.

\section*{Acknowledgements}
  The authors would like to thank Lucca Rawlyk, Fernanda Gubert, and Erik Almeida for their valuable help in this work. This study was partially supported by the project CNPq-URBCOMP (process 403260/2016-7), CAPES, CNPq, and Fundacao Araucaria. 
  

\bibliographystyle{bmc-mathphys} 


\newcommand{\BMCxmlcomment}[1]{}

\BMCxmlcomment{

<refgrp>

<bibl id="B1">
  <title><p>Building strategic relationships: How to extend your organization's
  reach through partnerships, alliances, and joint ventures</p></title>
  <aug>
    <au><snm>Bergquist</snm><fnm>WH</fnm></au>
    <au><snm>Betwee</snm><fnm>J</fnm></au>
    <au><snm>Meuel</snm><fnm>D</fnm></au>
  </aug>
  <publisher>San Franscisco, USA: Jossey-Bass Publishers</publisher>
  <pubdate>1995</pubdate>
</bibl>

<bibl id="B2">
  <title><p>An overview of strategic alliances</p></title>
  <aug>
    <au><snm>Elmuti</snm><fnm>D</fnm></au>
    <au><snm>Kathawala</snm><fnm>Y</fnm></au>
  </aug>
  <source>Management decision</source>
  <publisher>MCB UP Ltd</publisher>
  <pubdate>2001</pubdate>
  <volume>39</volume>
  <issue>3</issue>
  <fpage>205</fpage>
  <lpage>-218</lpage>
</bibl>

<bibl id="B3">
  <title><p>Can you measure the ROI of your social media marketing?</p></title>
  <aug>
    <au><snm>Hoffman</snm><fnm>DL</fnm></au>
    <au><snm>Fodor</snm><fnm>M</fnm></au>
  </aug>
  <source>MIT Sloan Management Review</source>
  <publisher>Massachusetts Institute of Technology, Cambridge, MA</publisher>
  <pubdate>2010</pubdate>
  <volume>52</volume>
  <issue>1</issue>
  <fpage>41</fpage>
</bibl>

<bibl id="B4">
  <title><p>Social media marketing</p></title>
  <aug>
    <au><snm>Tuten</snm><fnm>TL</fnm></au>
    <au><snm>Solomon</snm><fnm>MR</fnm></au>
  </aug>
  <publisher>Thousand Oaks, CA, USA: Sage</publisher>
  <pubdate>2017</pubdate>
</bibl>

<bibl id="B5">
  <title><p>Content marketing and brand engagement on social media: a study of
  Facebook{\'{}} s posts in the ecommerce industry in Brazil</p></title>
  <aug>
    <au><snm>Ferrari</snm><fnm>VC</fnm></au>
  </aug>
  <source>PhD thesis</source>
  <publisher>FGV - Fundação Getúlio Vargas</publisher>
  <pubdate>2016</pubdate>
</bibl>

<bibl id="B6">
  <title><p>Global social media research summary 2016</p></title>
  <aug>
    <au><snm>Chaffey</snm><fnm>D</fnm></au>
  </aug>
  <source>Smart Insights: Social Media Marketing</source>
  <pubdate>2016</pubdate>
</bibl>

<bibl id="B7">
  <title><p>Mining brand perceptions from Twitter social networks</p></title>
  <aug>
    <au><snm>Culotta</snm><fnm>A</fnm></au>
    <au><snm>Cutler</snm><fnm>J</fnm></au>
  </aug>
  <source>Marketing science</source>
  <publisher>INFORMS</publisher>
  <pubdate>2016</pubdate>
  <volume>35</volume>
  <issue>3</issue>
  <fpage>343</fpage>
  <lpage>-362</lpage>
</bibl>

<bibl id="B8">
  <title><p>Social media technology usage and customer relationship
  performance: A capabilities-based examination of social CRM</p></title>
  <aug>
    <au><snm>Trainor</snm><fnm>KJ</fnm></au>
    <au><snm>Andzulis</snm><fnm>JM</fnm></au>
    <au><snm>Rapp</snm><fnm>A</fnm></au>
    <au><snm>Agnihotri</snm><fnm>R</fnm></au>
  </aug>
  <source>Journal of Business Research</source>
  <publisher>Elsevier</publisher>
  <pubdate>2014</pubdate>
  <volume>67</volume>
  <issue>6</issue>
  <fpage>1201</fpage>
  <lpage>-1208</lpage>
</bibl>

<bibl id="B9">
  <title><p>Social media: Influencing customer satisfaction in B2B
  sales</p></title>
  <aug>
    <au><snm>Agnihotri</snm><fnm>R</fnm></au>
    <au><snm>Dingus</snm><fnm>R</fnm></au>
    <au><snm>Hu</snm><fnm>MY</fnm></au>
    <au><snm>Krush</snm><fnm>MT</fnm></au>
  </aug>
  <source>Industrial Marketing Management</source>
  <publisher>Elsevier</publisher>
  <pubdate>2016</pubdate>
  <volume>53</volume>
  <fpage>172</fpage>
  <lpage>-180</lpage>
</bibl>

<bibl id="B10">
  <title><p>The impact of social media on the consumer decision process:
  Implications for tourism marketing</p></title>
  <aug>
    <au><snm>Hudson</snm><fnm>S</fnm></au>
    <au><snm>Thal</snm><fnm>K</fnm></au>
  </aug>
  <source>Journal of Travel \& Tourism Marketing</source>
  <publisher>Taylor \& Francis</publisher>
  <pubdate>2013</pubdate>
  <volume>30</volume>
  <issue>1-2</issue>
  <fpage>156</fpage>
  <lpage>-160</lpage>
</bibl>

<bibl id="B11">
  <title><p>The livehoods project: Utilizing social media to understand the
  dynamics of a city</p></title>
  <aug>
    <au><snm>Cranshaw</snm><fnm>J</fnm></au>
    <au><snm>Schwartz</snm><fnm>R</fnm></au>
    <au><snm>Hong</snm><fnm>JI</fnm></au>
    <au><snm>Sadeh</snm><fnm>N</fnm></au>
  </aug>
  <source>Proc.\ of ICWSM'12</source>
  <publisher>Dublin, Ireland</publisher>
  <pubdate>2012</pubdate>
</bibl>

<bibl id="B12">
  <title><p>You Are What You Eat (and Drink): Identifying Cultural Boundaries
  by Analyzing Food and Drink Habits in Foursquare.</p></title>
  <aug>
    <au><snm>Silva</snm><fnm>TH</fnm></au>
    <au><snm>Melo</snm><fnm>POV</fnm></au>
    <au><snm>Almeida</snm><fnm>JM</fnm></au>
    <au><snm>Musolesi</snm><fnm>M</fnm></au>
    <au><snm>Loureiro</snm><fnm>AA</fnm></au>
  </aug>
  <source>Proc.\ of ICWSM'14</source>
  <publisher>Ann Arbor, USA</publisher>
  <pubdate>2014</pubdate>
</bibl>

<bibl id="B13">
  <title><p>Gender matters! Analyzing global cultural gender preferences for
  venues using social sensing</p></title>
  <aug>
    <au><snm>Mueller</snm><fnm>W</fnm></au>
    <au><snm>Silva</snm><fnm>TH</fnm></au>
    <au><snm>Almeida</snm><fnm>JM</fnm></au>
    <au><snm>Loureiro</snm><fnm>AA</fnm></au>
  </aug>
  <source>EPJ Data Science</source>
  <pubdate>2017</pubdate>
  <volume>6</volume>
  <issue>1</issue>
  <fpage>5</fpage>
  <url>http://dx.doi.org/10.1140/epjds/s13688-017-0101-0</url>
</bibl>

<bibl id="B14">
  <title><p>Cheers to Untappd! Preferences for Beer Reflect Cultural
  Differences Around the World</p></title>
  <aug>
    <au><snm>Brito</snm><fnm>S</fnm></au>
    <au><snm>Baldykowski</snm><fnm>A</fnm></au>
    <au><snm>Miczevski</snm><fnm>S</fnm></au>
    <au><snm>Silva</snm><fnm>TH</fnm></au>
  </aug>
  <source>Proc.\ of AMCIS'18</source>
  <publisher>New Orleans, USA</publisher>
  <pubdate>2018</pubdate>
</bibl>

<bibl id="B15">
  <title><p>Community detection in graphs</p></title>
  <aug>
    <au><snm>Fortunato</snm><fnm>S</fnm></au>
  </aug>
  <source>Physics reports</source>
  <publisher>Elsevier</publisher>
  <pubdate>2010</pubdate>
  <volume>486</volume>
  <issue>3-5</issue>
  <fpage>75</fpage>
  <lpage>-174</lpage>
</bibl>

<bibl id="B16">
  <title><p>Community discovery in dynamic networks: a survey</p></title>
  <aug>
    <au><snm>Rossetti</snm><fnm>G</fnm></au>
    <au><snm>Cazabet</snm><fnm>R</fnm></au>
  </aug>
  <source>ACM Computing Surveys</source>
  <publisher>ACM</publisher>
  <pubdate>2018</pubdate>
  <volume>51</volume>
  <issue>2</issue>
  <fpage>35</fpage>
</bibl>

<bibl id="B17">
  <title><p>Opinion mining in management research: the state of the art and the
  way forward</p></title>
  <aug>
    <au><snm>Mukhopadhyay</snm><fnm>S</fnm></au>
  </aug>
  <source>OPSEARCH</source>
  <publisher>Springer</publisher>
  <pubdate>2018</pubdate>
  <volume>55</volume>
  <issue>2</issue>
  <fpage>221</fpage>
  <lpage>-250</lpage>
</bibl>

<bibl id="B18">
  <title><p>Urban Computing Leveraging Location-Based Social Network Data: a
  Survey</p></title>
  <aug>
    <au><snm>Silva</snm><fnm>T</fnm></au>
    <au><snm>Viana</snm><fnm>A</fnm></au>
    <au><snm>Benevenuto</snm><fnm>F</fnm></au>
    <au><snm>Villas</snm><fnm>L</fnm></au>
    <au><snm>Salles</snm><fnm>J</fnm></au>
    <au><snm>Loureiro</snm><fnm>A</fnm></au>
    <au><snm>Queercia</snm><fnm>D</fnm></au>
  </aug>
  <source>ACM Computing Surveys</source>
  <pubdate>2019</pubdate>
  <volume>52</volume>
  <issue>1</issue>
  <fpage>17:1</fpage>
  <lpage>-17:39</lpage>
</bibl>

<bibl id="B19">
  <title><p>Beyond Sights: Large Scale Study of Tourists' Behavior Using
  Foursquare Data</p></title>
  <aug>
    <au><snm>Ferreira</snm><fnm>A. P. G.</fnm></au>
    <au><snm>Silva</snm><fnm>T. H.</fnm></au>
    <au><snm>Loureiro</snm><fnm>A. A. F.</fnm></au>
  </aug>
  <source>Proc.\ of IEEE ICDMW'15 Workshops</source>
  <pubdate>2015</pubdate>
  <fpage>1117</fpage>
  <lpage>1124</lpage>
</bibl>

<bibl id="B20">
  <title><p>Characterizing Geographic Variation in Well-Being Using
  Tweets.</p></title>
  <aug>
    <au><snm>Schwartz</snm><fnm>HA</fnm></au>
    <au><snm>Eichstaedt</snm><fnm>JC</fnm></au>
    <au><snm>Kern</snm><fnm>ML</fnm></au>
    <au><snm>Dziurzynski</snm><fnm>L</fnm></au>
    <au><snm>Lucas</snm><fnm>RE</fnm></au>
    <au><snm>Agrawal</snm><fnm>M</fnm></au>
    <au><snm>Park</snm><fnm>GJ</fnm></au>
    <au><snm>Lakshmikanth</snm><fnm>SK</fnm></au>
    <au><snm>Jha</snm><fnm>S</fnm></au>
    <au><snm>Seligman</snm><fnm>ME</fnm></au>
    <au><cnm>others</cnm></au>
  </aug>
  <source>Proc.\ of ICWSM'13</source>
  <publisher>Boston, USA</publisher>
  <pubdate>2013</pubdate>
</bibl>

<bibl id="B21">
  <title><p>Characterizing Dietary Choices, Nutrition, and Language in Food
  Deserts via Social Media</p></title>
  <aug>
    <au><snm>De Choudhury</snm><fnm>M</fnm></au>
    <au><snm>Sharma</snm><fnm>S</fnm></au>
    <au><snm>Kiciman</snm><fnm>E</fnm></au>
  </aug>
  <source>Proc.\ of CSCW'16</source>
  <publisher>San Francisco, USA: ACM</publisher>
  <pubdate>2016</pubdate>
  <fpage>1157</fpage>
  <lpage>-1170</lpage>
</bibl>

<bibl id="B22">
  <title><p>Chatty maps: constructing sound maps of urban areas from social
  media data</p></title>
  <aug>
    <au><snm>Aiello</snm><fnm>LM</fnm></au>
    <au><snm>Schifanella</snm><fnm>R</fnm></au>
    <au><snm>Quercia</snm><fnm>D</fnm></au>
    <au><snm>Aletta</snm><fnm>F</fnm></au>
  </aug>
  <source>Open Science</source>
  <publisher>The Royal Society</publisher>
  <pubdate>2016</pubdate>
  <volume>3</volume>
  <issue>3</issue>
  <fpage>150690</fpage>
</bibl>

<bibl id="B23">
  <title><p>{Uncovering the Perception of Urban Outdoor Areas Expressed in
  Social Media}</p></title>
  <aug>
    <au><snm>Santos</snm><fnm>FA</fnm></au>
    <au><snm>Silva</snm><fnm>TH</fnm></au>
    <au></au>
    <au><snm>Loureiro</snm><fnm>AAF</fnm></au>
    <au><snm>Villas</snm><fnm>LA</fnm></au>
  </aug>
  <source>Proc. of IEEE ACM Web Intelligence (WI)</source>
  <publisher>Santiago, Chile</publisher>
  <pubdate>2018</pubdate>
</bibl>

<bibl id="B24">
  <title><p>International Gender Differences and Gaps in Online Social
  Networks</p></title>
  <aug>
    <au><snm>Magno</snm><fnm>G</fnm></au>
    <au><snm>Weber</snm><fnm>I</fnm></au>
  </aug>
  <source>Proc.\ of SocInfo</source>
  <publisher>Barcelona, Spain: Springer</publisher>
  <pubdate>2014</pubdate>
  <fpage>121</fpage>
  <lpage>-138</lpage>
</bibl>

<bibl id="B25">
  <title><p>Exploring millions of footprints in location sharing
  services.</p></title>
  <aug>
    <au><snm>Cheng</snm><fnm>Z</fnm></au>
    <au><snm>Caverlee</snm><fnm>J</fnm></au>
    <au><snm>Lee</snm><fnm>K</fnm></au>
    <au><snm>Sui</snm><fnm>DZ</fnm></au>
  </aug>
  <source>Proc.\ of ICWSM'11</source>
  <publisher>Barcelona, Spain</publisher>
  <pubdate>2011</pubdate>
  <fpage>81</fpage>
  <lpage>-88</lpage>
</bibl>

<bibl id="B26">
  <title><p>The national geographic characteristics of online public opinion
  propagation in China based on WeChat network</p></title>
  <aug>
    <au><snm>Ai</snm><fnm>C</fnm></au>
    <au><snm>Chen</snm><fnm>B</fnm></au>
    <au><snm>He</snm><fnm>L</fnm></au>
    <au><snm>Lai</snm><fnm>K</fnm></au>
    <au><snm>Qiu</snm><fnm>X</fnm></au>
  </aug>
  <source>GeoInformatica</source>
  <publisher>Springer</publisher>
  <pubdate>2018</pubdate>
  <fpage>1</fpage>
  <lpage>-24</lpage>
</bibl>

<bibl id="B27">
  <title><p>Understanding online groups through social media</p></title>
  <aug>
    <au><snm>Barbier</snm><fnm>G</fnm></au>
    <au><snm>Tang</snm><fnm>L</fnm></au>
    <au><snm>Liu</snm><fnm>H</fnm></au>
  </aug>
  <source>Wiley Interdisciplinary Reviews: Data Mining and Knowledge
  Discovery</source>
  <publisher>Wiley Online Library</publisher>
  <pubdate>2011</pubdate>
  <volume>1</volume>
  <issue>4</issue>
  <fpage>330</fpage>
  <lpage>-338</lpage>
</bibl>

<bibl id="B28">
  <title><p>Structural correlation between communities and core-periphery
  structures in social networks: Evidence from Twitter data</p></title>
  <aug>
    <au><snm>Yang</snm><fnm>J</fnm></au>
    <au><snm>Zhang</snm><fnm>M</fnm></au>
    <au><snm>Shen</snm><fnm>KN</fnm></au>
    <au><snm>Ju</snm><fnm>X</fnm></au>
    <au><snm>Guo</snm><fnm>X</fnm></au>
  </aug>
  <source>Expert Systems with Applications</source>
  <pubdate>2018</pubdate>
  <volume>111</volume>
  <fpage>91</fpage>
  <lpage>99</lpage>
  <note>Big Data Analytics for Business Intelligence</note>
</bibl>

<bibl id="B29">
  <title><p>SIMPLE: a simplifying-ensembling framework for parallel community
  detection from large networks</p></title>
  <aug>
    <au><snm>Wu</snm><fnm>Z</fnm></au>
    <au><snm>Gao</snm><fnm>G</fnm></au>
    <au><snm>Bu</snm><fnm>Z</fnm></au>
    <au><snm>Cao</snm><fnm>J</fnm></au>
  </aug>
  <source>Cluster Computing</source>
  <publisher>Springer</publisher>
  <pubdate>2016</pubdate>
  <volume>19</volume>
  <issue>1</issue>
  <fpage>211</fpage>
  <lpage>-221</lpage>
</bibl>

<bibl id="B30">
  <title><p>Markov-network based latent link analysis for community detection
  in social behavioral interactions</p></title>
  <aug>
    <au><snm>Liu</snm><fnm>W</fnm></au>
    <au><snm>Yue</snm><fnm>K</fnm></au>
    <au><snm>Wu</snm><fnm>H</fnm></au>
    <au><snm>Fu</snm><fnm>X</fnm></au>
    <au><snm>Zhang</snm><fnm>Z</fnm></au>
    <au><snm>Huang</snm><fnm>W</fnm></au>
  </aug>
  <source>Applied Intelligence</source>
  <publisher>Springer</publisher>
  <pubdate>2017</pubdate>
  <fpage>1</fpage>
  <lpage>-16</lpage>
</bibl>

<bibl id="B31">
  <title><p>Social network data analytics for market segmentation in Indonesian
  telecommunications industry</p></title>
  <aug>
    <au><snm>Alamsyah</snm><fnm>A</fnm></au>
    <au><cnm>others</cnm></au>
  </aug>
  <source>Proc.\ of ICoICT'17</source>
  <pubdate>2017</pubdate>
  <fpage>1</fpage>
  <lpage>-5</lpage>
</bibl>

<bibl id="B32">
  <title><p>Understanding group structures and properties in social
  media</p></title>
  <aug>
    <au><snm>Tang</snm><fnm>L</fnm></au>
    <au><snm>Liu</snm><fnm>H</fnm></au>
  </aug>
  <source>Link Mining: Models, Algorithms, and Applications</source>
  <publisher>New York, NY, USA: Springer</publisher>
  <pubdate>2010</pubdate>
  <fpage>163</fpage>
  <lpage>-185</lpage>
</bibl>

<bibl id="B33">
  <title><p>Social Media Impact on Business Evaluation</p></title>
  <aug>
    <au><snm>Grizane</snm><fnm>T</fnm></au>
    <au><snm>Jurgelane</snm><fnm>I</fnm></au>
  </aug>
  <source>Procedia Computer Science</source>
  <publisher>Elsevier</publisher>
  <pubdate>2017</pubdate>
  <volume>104</volume>
  <fpage>190</fpage>
  <lpage>-196</lpage>
</bibl>

<bibl id="B34">
  <title><p>If we post it they will come: A small business perspective of
  social media marketing</p></title>
  <aug>
    <au><snm>Mahony</snm><fnm>T</fnm></au>
    <au><snm>Myers</snm><fnm>T</fnm></au>
    <au><snm>Low</snm><fnm>D</fnm></au>
    <au><snm>Eagle</snm><fnm>L</fnm></au>
  </aug>
  <source>Proc.\ of ACSW'18</source>
  <publisher>Brisbane, Australia</publisher>
  <pubdate>2018</pubdate>
  <fpage>21</fpage>
</bibl>

<bibl id="B35">
  <title><p>Exploiting time series analysis in Twitter to measure a campaign
  process performance</p></title>
  <aug>
    <au><snm>Kafeza</snm><fnm>E</fnm></au>
    <au><snm>Makris</snm><fnm>C</fnm></au>
    <au><snm>Rompolas</snm><fnm>G</fnm></au>
  </aug>
  <source>Proc.\ of SCC'17</source>
  <pubdate>2017</pubdate>
  <fpage>68</fpage>
  <lpage>-75</lpage>
</bibl>

<bibl id="B36">
  <title><p>User communities evolution in microblogs: A public awareness
  barometer for real world events</p></title>
  <aug>
    <au><snm>Giatsoglou</snm><fnm>M</fnm></au>
    <au><snm>Chatzakou</snm><fnm>D</fnm></au>
    <au><snm>Vakali</snm><fnm>A</fnm></au>
  </aug>
  <source>World Wide Web</source>
  <publisher>Springer</publisher>
  <pubdate>2015</pubdate>
  <volume>18</volume>
  <issue>5</issue>
  <fpage>1269</fpage>
  <lpage>-1299</lpage>
</bibl>

<bibl id="B37">
  <title><p>Visual analytics for exploring topic long-term evolution and
  detecting weak signals in company targeted tweets</p></title>
  <aug>
    <au><snm>Pepin</snm><fnm>L</fnm></au>
    <au><snm>Kuntz</snm><fnm>P</fnm></au>
    <au><snm>Blanchard</snm><fnm>J</fnm></au>
    <au><snm>Guillet</snm><fnm>F</fnm></au>
    <au><snm>Suignard</snm><fnm>P</fnm></au>
  </aug>
  <source>Computers \& Industrial Engineering</source>
  <publisher>Elsevier</publisher>
  <pubdate>2017</pubdate>
  <volume>112</volume>
  <fpage>450</fpage>
  <lpage>-458</lpage>
</bibl>

<bibl id="B38">
  <title><p>Excavating social circles via user interests</p></title>
  <aug>
    <au><snm>Palsetia</snm><fnm>D</fnm></au>
    <au><snm>Patwary</snm><fnm>MMA</fnm></au>
    <au><snm>Agrawal</snm><fnm>A</fnm></au>
    <au><snm>Choudhary</snm><fnm>A</fnm></au>
  </aug>
  <source>Social Network Analysis and Mining</source>
  <publisher>Springer</publisher>
  <pubdate>2014</pubdate>
  <volume>4</volume>
  <issue>1</issue>
  <fpage>170</fpage>
</bibl>

<bibl id="B39">
  <title><p>Where is the Goldmine?: Finding Promising Business Locations
  through Facebook Data Analytics</p></title>
  <aug>
    <au><snm>Lin</snm><fnm>J</fnm></au>
    <au><snm>Oentaryo</snm><fnm>R</fnm></au>
    <au><snm>Lim</snm><fnm>EP</fnm></au>
    <au><snm>Vu</snm><fnm>C</fnm></au>
    <au><snm>Vu</snm><fnm>A</fnm></au>
    <au><snm>Kwee</snm><fnm>A</fnm></au>
  </aug>
  <source>Proc.\ of Hypertext'16</source>
  <publisher>Halifax, Canada</publisher>
  <pubdate>2016</pubdate>
  <fpage>93</fpage>
  <lpage>-102</lpage>
</bibl>

<bibl id="B40">
  <title><p>Geo-spotting: Mining Online Location-based Services for Optimal
  Retail Store Placement</p></title>
  <aug>
    <au><snm>Karamshuk</snm><fnm>D</fnm></au>
    <au><snm>Noulas</snm><fnm>A</fnm></au>
    <au><snm>Scellato</snm><fnm>S</fnm></au>
    <au><snm>Nicosia</snm><fnm>V</fnm></au>
    <au><snm>Mascolo</snm><fnm>C</fnm></au>
  </aug>
  <source>Proc.\ of ACM KDD'13</source>
  <publisher>Chicago, Illinois, USA</publisher>
  <pubdate>2013</pubdate>
  <fpage>793</fpage>
  <lpage>-801</lpage>
  <url>http://doi.acm.org/10.1145/2487575.2487616</url>
</bibl>

<bibl id="B41">
  <title><p>Identificando a Relação Virtual Entre Empresas Explorando
  Reações de Usuários no Facebook</p></title>
  <aug>
    <au><snm>Tsutsumi</snm><fnm>D</fnm></au>
    <au><snm>Fenerich</snm><fnm>A</fnm></au>
    <au><snm>Silva</snm><fnm>TH</fnm></au>
  </aug>
  <source>Proc.\ of CoUrb'18</source>
  <publisher>Campos do Jordão, Brazil</publisher>
  <pubdate>2018</pubdate>
</bibl>

<bibl id="B42">
  <title><p>State of the Geotags: Motivations and Recent Changes.</p></title>
  <aug>
    <au><snm>Tasse</snm><fnm>D</fnm></au>
    <au><snm>Liu</snm><fnm>Z</fnm></au>
    <au><snm>Sciuto</snm><fnm>A</fnm></au>
    <au><snm>Hong</snm><fnm>JI</fnm></au>
  </aug>
  <source>Proc.\ of ICWSM'17</source>
  <publisher>Montreal, Canadá</publisher>
  <pubdate>2017</pubdate>
  <fpage>250</fpage>
  <lpage>-259</lpage>
</bibl>

<bibl id="B43">
  <title><p>Near linear time algorithm to detect community structures in
  large-scale networks</p></title>
  <aug>
    <au><snm>Raghavan</snm><fnm>UN</fnm></au>
    <au><snm>Albert</snm><fnm>R</fnm></au>
    <au><snm>Kumara</snm><fnm>S</fnm></au>
  </aug>
  <source>Physical review E</source>
  <publisher>APS</publisher>
  <pubdate>2007</pubdate>
  <volume>76</volume>
  <issue>3</issue>
  <fpage>036106</fpage>
</bibl>

<bibl id="B44">
  <title><p>Clustering algorithms</p></title>
  <aug>
    <au><snm>Hartigan</snm><fnm>JA</fnm></au>
  </aug>
  <publisher>Wiley</publisher>
  <pubdate>1975</pubdate>
</bibl>

</refgrp>
} 

\end{backmatter}

\clearpage

\begin{appendices}
\section{Supplementary information on business reaction database and business relationship graph}\label{app:reacDatabase}

\begin{figure}[ht]
    \includegraphics[width=.85\textwidth]{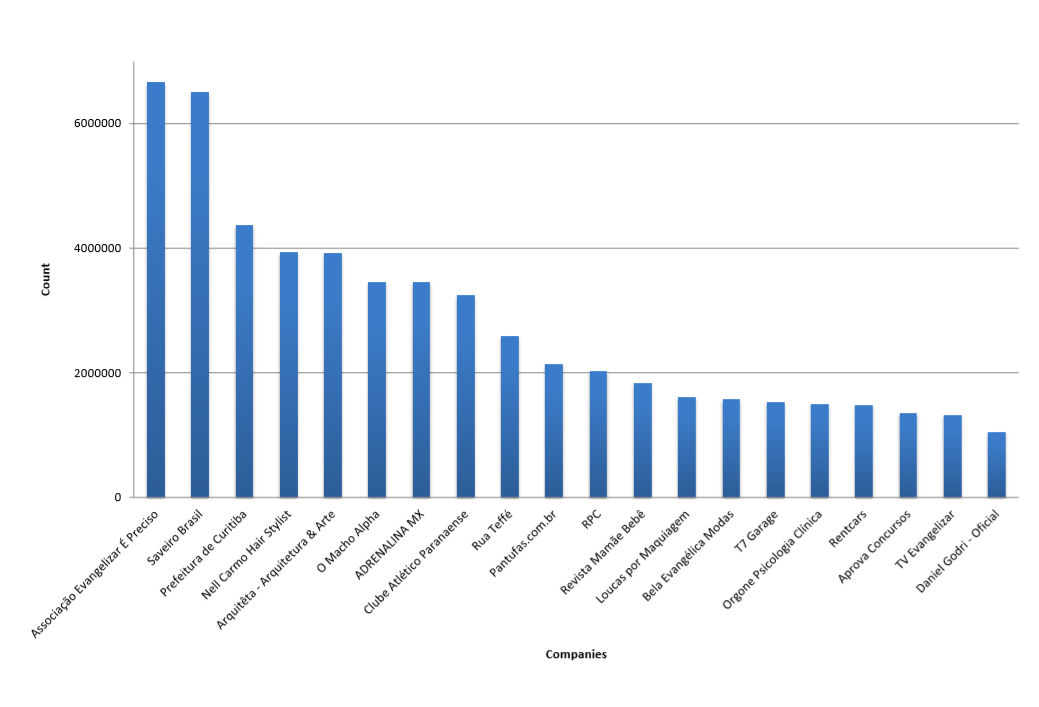}
    \caption{Histogram of the top twenty companies in terms of user reaction number. }
    \label{fig:compLike}
\end{figure}

\begin{figure}[ht]
    \includegraphics[width=.85\textwidth]{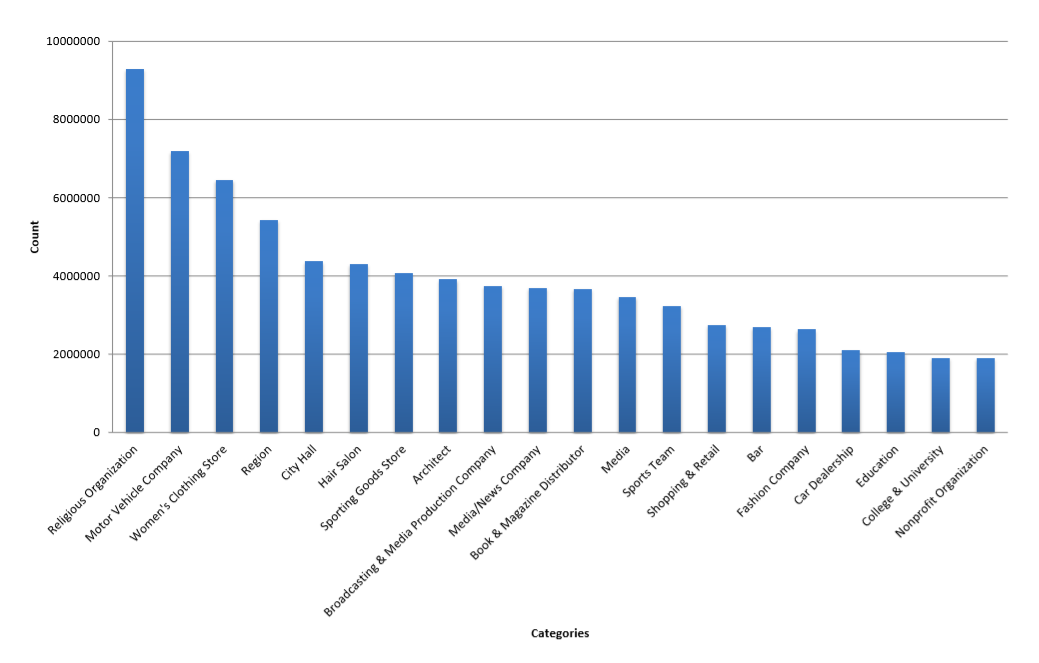}
    \caption{Histogram of the top twenty categories in terms of user reaction number. The names are the original ones and do not reflect the renaming process shown in figure \ref{fig:FaceCategories}.}
    \label{fig:catLike}
\end{figure}

\begin{figure}[ht]
    \includegraphics[width=.55\textwidth]{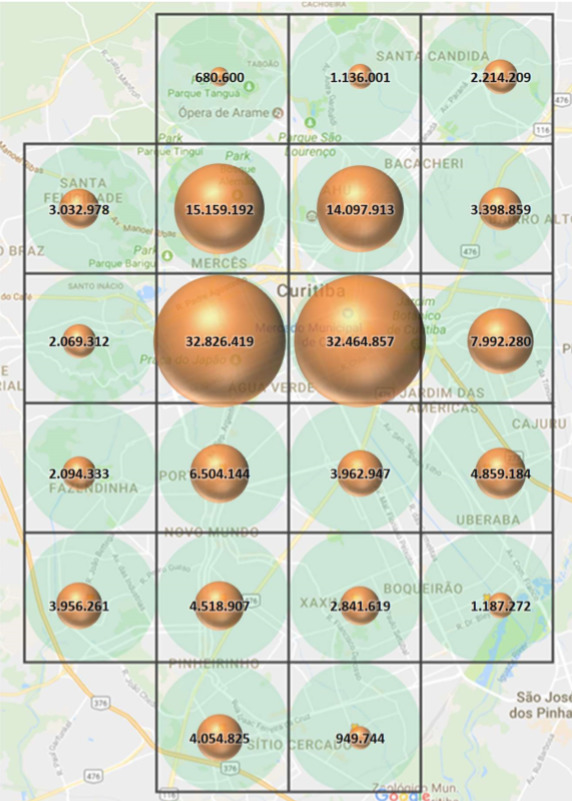}
    \caption{A map showing the amount of user reactions per region considered in the collection process presented in Section \ref{sec:datacollection}.}
    \label{fig:mapLike}
\end{figure}

\begin{table}[ht]
\caption{Business graph nodes ranked by number of connections.}
\label{tab:graphNodeRanking}
\tiny
\begin{tabular}{lll}
\hline
\rowcolor[HTML]{EFEFEF} 
\textbf{Ranking} & \textbf{Business Name}                    & \textbf{Number of Connected Edges} \\ \hline
1                & Prefeitura de Curitiba                    & 1396                               \\
\rowcolor[HTML]{EFEFEF} 
2                & RPC                                       & 1357                               \\
3                & Portal Banda B                            & 1329                               \\
\rowcolor[HTML]{EFEFEF} 
4                & Clube Atlético Paranaense                 & 1273                               \\
5                & Aliança Móveis                            & 1264                               \\
\rowcolor[HTML]{EFEFEF} 
6                & Shopping Mueller                          & 1236                               \\
7                & Dr. Freeze                                & 1200                               \\
\rowcolor[HTML]{EFEFEF} 
8                & CWB Brasil                                & 1200                               \\
9                & Curitiba Comedy Club                      & 1199                               \\
\rowcolor[HTML]{EFEFEF} 
10               & Shopping Palladium - Curitiba             & 1196                               \\
11               & ParkShopping Barigüi                      & 1196                               \\
\rowcolor[HTML]{EFEFEF} 
12               & Disk Ingressos                            & 1164                               \\
13               & Ripa na Chulipa                           & 1148                               \\
\rowcolor[HTML]{EFEFEF} 
14               & Taco el Pancho                            & 1148                               \\
15               & Integra - Cursos \& Treinamentos Curitiba & 1138                               \\
\rowcolor[HTML]{EFEFEF} 
16               & Saia Justa Vestidos                       & 1132                               \\
17               & Cosmeticos Curitiba Centro                & 1100                               \\
\rowcolor[HTML]{EFEFEF} 
18               & Leve Sabor                                & 1100                               \\
19               & Zapata  Bar                               & 1099                               \\
\rowcolor[HTML]{EFEFEF} 
20               & Universidade Tuiuti do Paraná - UTP       & 1093                               \\
21               & Espaço Gourmet Escola de Gastronomia      & 1086                               \\
\rowcolor[HTML]{EFEFEF} 
22               & Pele Morena Lingerie                      & 1086                               \\
23               & Invisible Braces Ortodontia               & 1083                               \\
\rowcolor[HTML]{EFEFEF} 
24               & Mais 55                                   & 1077                               \\
25               & Daju                                      & 1077                               \\
\rowcolor[HTML]{EFEFEF} 
26               & La Passion Palladium                      & 1074                               \\
27               & Vida Leve Refeições Saudáveis             & 1071                               \\
\rowcolor[HTML]{EFEFEF} 
28               & Vivá                                      & 1063                               \\
29               & Blood Rock Bar                            & 1061                               \\
\rowcolor[HTML]{EFEFEF} 
30               & Rádio 98FM Curitiba                       & 1052                               \\
31               & Bar Quermesse                             & 1047                               \\
\rowcolor[HTML]{EFEFEF} 
32               & Seven Entretenimento                      & 1047                               \\
33               & Power Airsoft                             & 1045                               \\
\rowcolor[HTML]{EFEFEF} 
34               & Bar e Restaurante Hora Extra              & 1035                               \\
35               & Hallorino Jr                              & 1032                               \\
\rowcolor[HTML]{EFEFEF} 
36               & PolloShop                                 & 1025                               \\
37               & Nell Carmo Hair Stylist                   & 1025                               \\
\rowcolor[HTML]{EFEFEF} 
38               & BandNews FM Curitiba                      & 1020                               \\
39               & Cantinho da Bica                          & 1010                               \\
\rowcolor[HTML]{EFEFEF} 
40               & Rádio Mundo Livre FM                      & 1000                               \\
41               & Império da Pizza                          & 995                                \\
\rowcolor[HTML]{EFEFEF} 
42               & DOCK Premium                              & 991                                \\
43               & Autoescola Bello                          & 991                                \\
\rowcolor[HTML]{EFEFEF} 
44               & A Barateira Feirão de Calçados            & 989                                \\
45               & Hospital Pequeno Príncipe                 & 988                                \\
\rowcolor[HTML]{EFEFEF} 
46               & Barone Consignações                       & 985                                \\
47               & Tatica Imoveis                            & 984                                \\
\rowcolor[HTML]{EFEFEF} 
48               & Shopping Total                            & 981                                \\
49               & Casa da Bruxa                             & 973                                \\
\rowcolor[HTML]{EFEFEF} 
50               & Plus Santé Emergências Médicas Ltda       & 973                               
\end{tabular}
\end{table}

\begin{table}[ht]
\caption{Business graph edges ranked by weight.}
\label{tab:graphEdgeRanking}
\tiny
\begin{tabular}{llll}
\hline
\rowcolor[HTML]{EFEFEF} 
\textbf{Ranking} & \textbf{Edge Node}                        & \textbf{Edge Node}                                & \multicolumn{1}{l}{\textbf{Edge Weight}} \\ \hline
1                & RPC                                       & Prefeitura de Curitiba                            & 127958                                    \\
\rowcolor[HTML]{EFEFEF} 
2                & Portal Banda B                            & Prefeitura de Curitiba                            & 82904                                     \\
3                & Loja Us Store                             & Vissothi                                          & 75629                                     \\
\rowcolor[HTML]{EFEFEF} 
4                & Portal Banda B                            & RPC                                               & 74084                                     \\
5                & Clube Atlético Paranaense                 & Prefeitura de Curitiba                            & 73948                                     \\
\rowcolor[HTML]{EFEFEF} 
6                & Associação Evangelizar É Preciso          & TV Evangelizar                                    & 72184                                     \\
7                & Loja Us Store                             & Pantufas.com.br                                   & 57844                                     \\
\rowcolor[HTML]{EFEFEF} 
8                & Clube Atlético Paranaense                 & RPC                                               & 56555                                     \\
9                & Shopping Mueller                          & Prefeitura de Curitiba                            & 54823                                     \\
\rowcolor[HTML]{EFEFEF} 
10               & Clube Atlético Paranaense                 & Portal Banda B                                    & 54437                                     \\
11               & ParkShopping Barigüi                      & Prefeitura de Curitiba                            & 51522                                     \\
\rowcolor[HTML]{EFEFEF} 
12               & Curitiba Cult                             & Prefeitura de Curitiba                            & 50463                                     \\
13               & Shopping Palladium - Curitiba             & Prefeitura de Curitiba                            & 49198                                     \\
\rowcolor[HTML]{EFEFEF} 
14               & BandNews FM Curitiba                      & Prefeitura de Curitiba                            & 46635                                     \\
15               & Universidade Tuiuti do Paraná - UTP       & Prefeitura de Curitiba                            & 45870                                     \\
\rowcolor[HTML]{EFEFEF} 
16               & Vissothi                                  & Pantufas.com.br                                   & 45213                                     \\
17               & Arquitêta - Arquitetura \& Arte           & Prefeitura de Curitiba                            & 45194                                     \\
\rowcolor[HTML]{EFEFEF} 
18               & Blood Rock Bar                            & Prefeitura de Curitiba                            & 44101                                     \\
19               & Rádio Mundo Livre FM                      & Prefeitura de Curitiba                            & 44033                                     \\
\rowcolor[HTML]{EFEFEF} 
20               & ParkShopping Barigüi                      & Shopping Mueller                                  & 43942                                     \\
21               & Saveiro Brasil                            & ADRENALINA MX                                     & 42377                                     \\
\rowcolor[HTML]{EFEFEF} 
22               & Dr. Freeze                                & Prefeitura de Curitiba                            & 41481                                     \\
23               & Ripa na Chulipa                           & Aliança Móveis                                    & 41188                                     \\
\rowcolor[HTML]{EFEFEF} 
24               & CWB Brasil                                & Prefeitura de Curitiba                            & 41041                                     \\
25               & Portal Banda B                            & Aliança Móveis                                    & 40995                                     \\
\rowcolor[HTML]{EFEFEF} 
26               & Bella Morena Boutique                     & Rua Teffé                                         & 40642                                     \\
27               & Shopping Mueller                          & Aliança Móveis                                    & 39563                                     \\
\rowcolor[HTML]{EFEFEF} 
28               & Pele Morena Lingerie                      & Aliança Móveis                                    & 38903                                     \\
29               & RPC                                       & Aliança Móveis                                    & 38812                                     \\
\rowcolor[HTML]{EFEFEF} 
30               & Prefeitura de Curitiba                    & Curitiba Comedy Club                              & 38638                                     \\
31               & Shopping Mueller                          & RPC                                               & 38469                                     \\
\rowcolor[HTML]{EFEFEF} 
32               & Prefeitura de Curitiba                    & Aliança Móveis                                    & 38298                                     \\
33               & Shopping Palladium - Curitiba             & RPC                                               & 38193                                     \\
\rowcolor[HTML]{EFEFEF} 
34               & Pantufas.com.br                           & Saveiro Brasil                                    & 38159                                     \\
35               & Hallorino Jr                              & RPC                                               & 37498                                     \\
\rowcolor[HTML]{EFEFEF} 
36               & ParkShopping Barigüi                      & RPC                                               & 37350                                     \\
37               & Hospital Pequeno Príncipe                 & Prefeitura de Curitiba                            & 36904                                     \\
\rowcolor[HTML]{EFEFEF} 
38               & Associação Evangelizar É Preciso          & Exper. de Deus com padre Reg. Manzotti & 36505                                     \\
39               & Disk Ingressos                            & Prefeitura de Curitiba                            & 36248                                     \\
\rowcolor[HTML]{EFEFEF} 
40               & Integra - Cursos \& Treinamentos Curitiba & Aliança Móveis                                    & 35842                                     \\
41               & Espaço Gourmet Escola de Gastronomia      & Prefeitura de Curitiba                            & 35641                                     \\
\rowcolor[HTML]{EFEFEF} 
42               & CWB Brasil                                & RPC                                               & 35437                                     \\
43               & RPC                                       & BandNews FM Curitiba                              & 35045                                     \\
\rowcolor[HTML]{EFEFEF} 
44               & Saia Justa Vestidos                       & Aliança Móveis                                    & 34837                                     \\
45               & Portal Banda B                            & CWB Brasil                                        & 34823                                     \\
\rowcolor[HTML]{EFEFEF} 
46               & Portal Banda B                            & Shopping Palladium - Curitiba                     & 34641                                     \\
47               & TV Evangelizar                            & Exper. de Deus com padre Reg. Manzotti & 34596                                     \\
\rowcolor[HTML]{EFEFEF} 
48               & Invisible Braces Ortodontia               & Aliança Móveis                                    & 33152                                     \\
49               & Aliança Móveis                            & Barone Consignações                               & 33150                                     \\
\rowcolor[HTML]{EFEFEF} 
50               & Portal Banda B                            & Shopping Mueller                                  & 32969                                    
\end{tabular}
\end{table}

\end{appendices}

\end{document}